\documentclass[a4paper,11pt]{article}
\pdfoutput=1

\usepackage{jheppub}

\usepackage{multirow}

\usepackage{subfig}
\usepackage{xspace}
\usepackage[countmax]{subfloat}
\usepackage{slashed}

\usepackage{comment}

\usepackage{bbm}

\newcommand{\nbar}{{\bar n}}

\newcommand{\sja}{n_{sj}}
\newcommand{\sjabar}{\bar{n}_{sj}}

\def\log{\text{log}}

\def\be{\begin{equation}}
\def\ee{\end{equation}}

\newcommand{\nn}{\nonumber}

\def\nbar{\bar n}

\DeclareRobustCommand{\Sec}[1]{Sec.~\ref{#1}}
\DeclareRobustCommand{\Secs}[2]{Secs.~\ref{#1} and \ref{#2}}
\DeclareRobustCommand{\App}[1]{App.~\ref{#1}}

\DeclareRobustCommand{\Fig}[1]{Fig.~\ref{#1}}
\DeclareRobustCommand{\Figs}[2]{Figs.~\ref{#1} and \ref{#2}}
\DeclareRobustCommand{\Eq}[1]{Eq.~(\ref{#1})}
\DeclareRobustCommand{\Eqs}[2]{Eqs.~(\ref{#1}) and (\ref{#2})}
\DeclareRobustCommand{\Ref}[1]{Ref.~\cite{#1}}
\DeclareRobustCommand{\Refs}[1]{Refs.~\cite{#1}}

\usepackage{color}

\preprint{ \begin{flushright}MIT--CTP 4821\\
LA-UR-16-26925
\end{flushright}
}

\title{The Analytic Structure of Non-Global Logarithms:\\
Convergence of the Dressed Gluon Expansion}

\author[1,2]{Andrew J. Larkoski,}
\affiliation[1]{Center for the Fundamental Laws of Nature, Harvard University, Cambridge, MA 02138, USA}
\affiliation[2]{Physics Department, Reed College, Portland, OR 97202,USA}

\author[3,4,5]{Ian Moult}
\affiliation[3]{Berkeley Center for Theoretical Physics, University of California, Berkeley, CA 94720, USA}
\affiliation[4]{Theoretical Physics Group, Lawrence Berkeley National Laboratory, Berkeley, CA 94720, USA}
\affiliation[5]{Center for Theoretical Physics, Massachusetts Institute of Technology, Cambridge, MA 02139, USA}
\author[5,6]{and Duff Neill}
\affiliation[6]{Theoretical Division, MS B283, Los Alamos National Laboratory, Los Alamos, NM 87545, USA}

\emailAdd{larkoski@reed.edu}
\emailAdd{ianmoult@lbl.gov}
\emailAdd{dneill@mit.edu}

\abstract{
Non-global logarithms (NGLs) are the leading manifestation of correlations between distinct phase space regions in QCD and gauge theories and have proven a challenge to understand using traditional resummation techniques.  Recently, the dressed gluon expansion was introduced that enables an expansion of the NGL series in terms of a ``dressed gluon" building block, defined by an all-orders factorization theorem.  Here, we clarify the nature of the dressed gluon expansion, and prove that it has an infinite radius of convergence as a solution to the leading logarithmic and large-$N_c$ master equation for NGLs, the Banfi-Marchesini-Smye (BMS) equation. The dressed gluon expansion therefore provides an expansion of the NGL series that can be truncated at any order, with reliable uncertainty estimates. In contrast, manifest in the results of the fixed-order expansion of the BMS equation up to 12-loops is a breakdown of convergence at a finite value of $\alpha_s$log. We explain this finite radius of convergence using the dressed gluon expansion, showing how the dynamics of the buffer region, a region of phase space near the boundary of the jet that was identified in early studies of NGLs, leads to large contributions to the fixed order expansion. We also use the dressed gluon expansion to discuss the convergence of the next-to-leading NGL series, and the role of collinear logarithms that appear at this order. Finally, we show how an understanding of the analytic behavior obtained from the dressed gluon expansion allows us to improve the fixed order NGL series using conformal transformations to extend the domain of analyticity. This allows us to calculate the NGL distribution for all values of $\alpha_s$log from the coefficients of the fixed order expansion.
}

\begin{document} 
\maketitle

\section{Introduction}\label{sec:intro}

Non-global logarithms (NGLs) \cite{Dasgupta:2001sh} have proven to be an obstruction to understanding the all-orders logarithmic structure of observables measured on jets or other restricted phase space regions. Since their discovery, there have been significant advances in their calculation \cite{Dasgupta:2002bw,Dasgupta:2002dc,Banfi:2002hw,Appleby:2002ke,Weigert:2003mm,Rubin:2010fc,Banfi:2010pa,Kelley:2011tj,Hornig:2011iu,Hornig:2011tg,Kelley:2011aa,Kelley:2012kj,Hatta:2013iba,Schwartz:2014wha,Khelifa-Kerfa:2015mma,Caron-Huot:2015bja,Larkoski:2015zka,Hagiwara:2015bia,Becher:2015hka,Neill:2015nya,Becher:2016mmh}, largely influenced by the development of the leading logarithmic (LL) and large-$N_c$ Banfi-Marchesini-Smye (BMS) equation \cite{Banfi:2002hw}. The dressed gluon expansion of \Ref{Larkoski:2015zka} proposed a method for reorganizing the degrees of freedom that contribute to NGLs into an expansion in identified soft jets, referred to as dressed gluons.\footnote{Similar ideas in the context of rapidity gaps were presented in \Refs{Forshaw:2006fk,Forshaw:2008cq,Forshaw:2009fz,DuranDelgado:2011tp}.} The dressed gluon is defined by an all-orders factorization theorem, whose associated resummation, dictated by renormalization group evolution, dresses the jet with an infinite number of unresolved gluons. By summing over dressed gluons, it was hoped that a convergent expansion of the NGL series could be obtained and that contained, at each order in the dressed gluon expansion, information about the all-orders behavior in the $\alpha_s$ expansion.

\Ref{Larkoski:2015zka} did not, however, formalize the nature of the dressed gluon expansion. There, the dressed gluon expansion was justified through numerical studies and comparison to the fixed-order expansion of the BMS equation; whether the series could be reliably truncated was never established, and its precise definition at higher orders was only guessed.
A reasonable criteria to judge an expansion of a series is that to reach any pre-defined accuracy requires only calculating a finite number of terms in the expansion. Because NGLs can be arbitrarily large, this requires the expansion to have an infinite radius of convergence if it is to describe the physics of the distribution in all regions. Indeed, due to exponentiation, this simple requirement is satisfied for the fixed-order expansion of familiar global logarithms (perhaps in some conjugate space), because the exponential function has an infinite radius of convergence.  For the dressed gluon expansion to be useful operationally, it should therefore have an infinite radius of convergence.

In this paper, we will prove that the dressed gluon expansion of the BMS equation for summing NGLs has an infinite radius of convergence, and we will clarify the nature of the expansion by relating the dressed gluon expansion to the method of successive approximations, a technique for solving ordinary differential equations (see, e.g., \Ref{CorduneanuDiff}) that produces convergent series as the solution. With a proof that the dressed gluon expansion absolutely converges to the full solution of the BMS equation, this provides a powerful analytic tool to investigate the structure of NGL series.  Additionally, because the dressed gluon expansion converges, the uncertainty introduced by truncating to a finite accuracy is well-defined.

While the proof of the convergence that we present is quite technical, the physical interpretation of the infinite radius of convergence is simple.  Higher orders in the dressed gluon expansion correspond to probing smaller infrared scales increasingly differentially.  For any fixed value of the NGL, including higher dressed gluons is necessary to span the gap from the high energy scale down to the measured scale of the NGL.  However, this is only true to a point: eventually, sufficiently many dressed gluons will be included to eliminate all large hierarchies.  For any finite value of the NGL this saturation occurs at a finite order in the dressed gluon expansion, and including higher order terms only refines the result.

A remarkable feature of the dressed gluon expansion is that when expanded in $\alpha_s$, the dressed gluon itself has a radius of convergence of $|L|\leq 1$, where for the particular case of NGLs in the heavy and light hemisphere masses $m_H$ and $m_L$:
\begin{equation}\label{eq:ldef}
L=\int\limits_{m_L}^{m_H} \frac{dm}{m} \frac{\alpha_s(m)}{\pi} N_c \simeq\frac{\alpha_s}{\pi} N_c \log\frac{m_H}{m_L}\,,
\end{equation}
where the rightmost expression is the result for a fixed-coupling. This behavior arises due to a singularity in the dressed gluon at $L=-1$ in the complex plane. Beyond $L=1$, in the region where the NGLs are large and truly necessitate resummation, the dressed gluon is capturing physics that cannot be reproduced by a fixed order expansion.\footnote{We emphasize that this is the radius of expansion in $L$, so that even expanding in $L$, but keeping running coupling effects does not help the radius of convergence.}  We show that this finite radius of convergence can be understood from the existence of the so called buffer region of the BMS equation \cite{Dasgupta:2002bw}, where radiation near the jet boundary is prohibited. We discuss in detail how the resummation associated with the dressed gluon expansion causes a breakdown in the perturbative expansion, in particular quantifying growing contributions at sub-leading logarithmic order. For the dressed gluon itself, no problems are found, since all such terms are automatically resummed.

With an understanding of the analytic structure of the dressed gluon, we will then argue that the fixed order $\alpha_s$ expansion of the leading NGL series also has a finite radius of convergence of $L=1$. This implies that a reorganization of the expansion, for example in terms of dressed gluons, is not only convenient, but necessary. We compare the known Monte Carlo resummation \cite{Dasgupta:2001sh} to the the explicit expansion of the BMS equation to 12-loop order, finding a barrier to continuation beyond $L=1$.\footnote{We thank Simon Caron-Huot for providing us with the 12-loop perturbative expansion.} Using our knowledge of the singularity at $L=-1$, we are able to apply a conformal transformation to reorganize the perturbative series, extending its domain of analyticity, and allowing us to calculate the NGL distribution for all values of $L$ from the coefficients of the fixed order expansion. We also use the dressed gluon expansion to comment on the behavior of the next-to-leading NGL series, in particular, focusing on the role of collinear logarithms.

The outline of this paper is as follows. We begin by briefly reviewing the physics of NGLs and the BMS equation in \Sec{sec:physics_BMS}, setting up our notation and language for the rest of the paper. In \Sec{sec:pic}, we review the method of successive approximations for solving differential equations and relate it to the dressed gluon expansion.
 Our proof that the dressed gluon expansion of the BMS equation converges is presented in \Sec{sec:dga}. 
 In \Sec{sec:fo_fail} we study the analytic structure of the dressed gluons, and show that when expanded in $\alpha_s$, they have a radius of convergence of $L=1$ due to the presence of the buffer region. By comparison to the results of the expansion of the BMS equation up to 12-loop order, we show that this breakdown of convergence is also manifest in the behavior of the perturbative series. Using an understanding of the analytic structure, obtained by studying the dressed gluon expansion, we also show how conformal mappings can be used to improve the behavior of the fixed order perturbative expansion. We conclude in \Sec{sec:conc}.

\section{Physics of Non-Global Logarithms and the BMS Equation}\label{sec:physics_BMS}

To set the stage for establishing the convergence of the dressed gluon expansion, we review the physics of NGLs and their leading-logarithmic and leading-color resummation as described by the BMS equation.  The BMS equation with full color is known to one- and two-loops (see \Refs{Weigert:2003mm,Caron-Huot:2015bja}), and in $\mathcal{N}=4$ Super Yang-Mills theory to three loops in the large-$N_c$ limit \Ref{Caron-Huot:2016tzz}. 

\begin{figure}
\begin{center}
\subfloat[]{\label{fig:GL_picture}
\includegraphics[width=6.75cm]{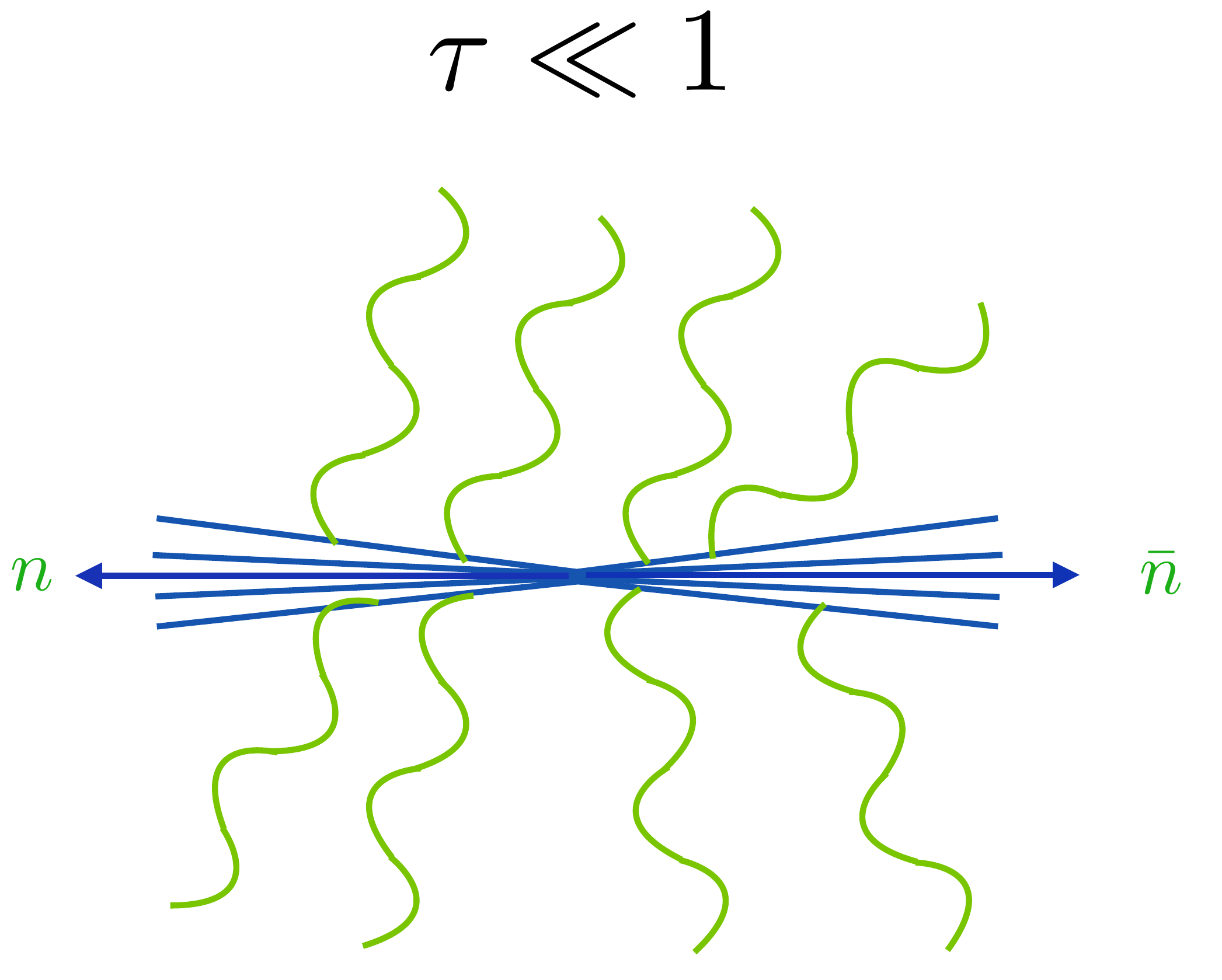}
}
\subfloat[]{\label{fig:NGL_picture}
\includegraphics[width=6.75cm]{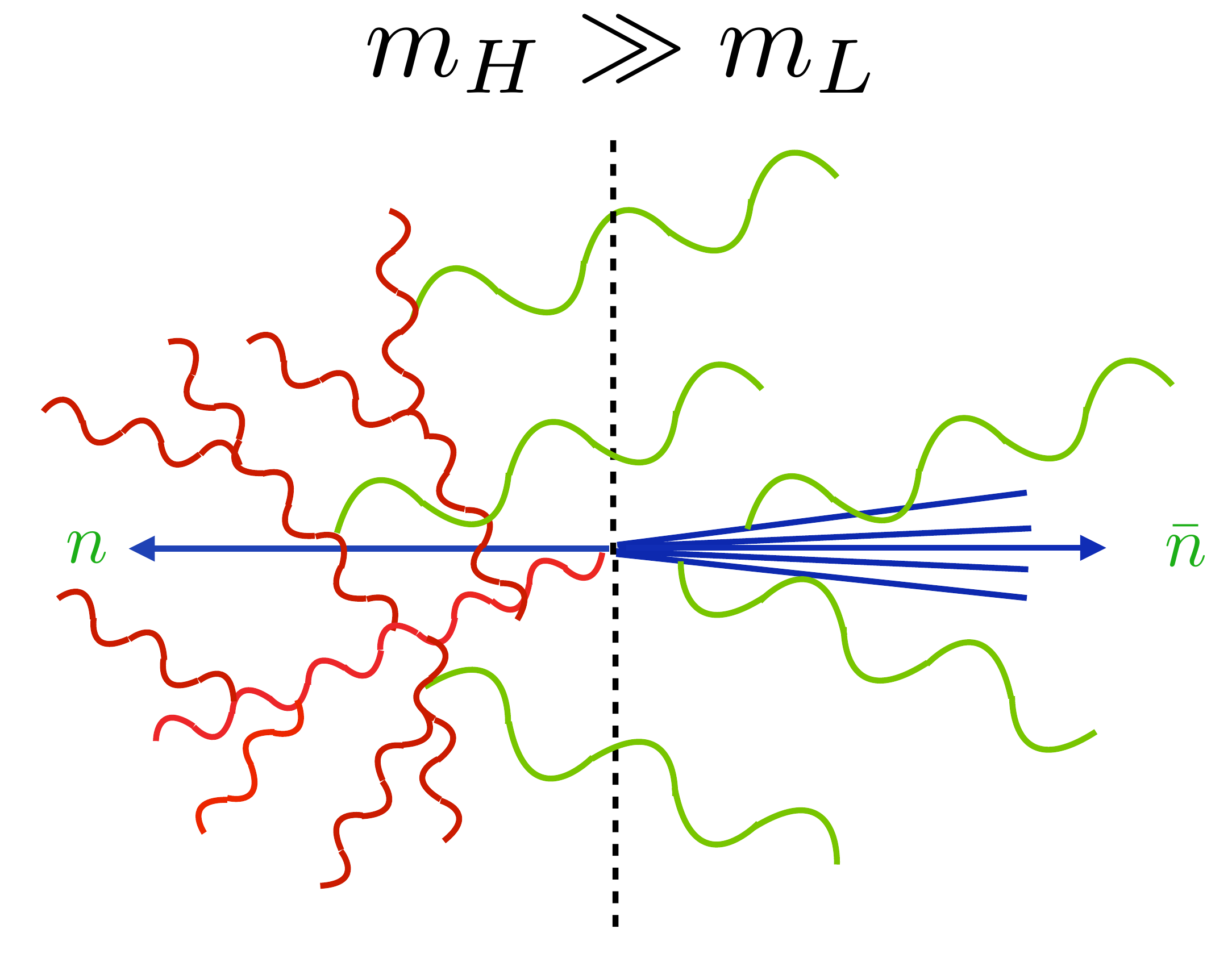}
}
\end{center}
\vspace{-0.2cm}
\caption{A schematic comparison of global and non-global logarithms. In (a) a global measurement is made, so that all real radiation is constrained by the measurement. No emissions occur between the scales, and all soft radiation is sourced by the eikonal lines in the directions of the original partons produced in the hard scattering.  In (b), a measurement is made in the right hemisphere. Real emissions in the left hemisphere are unconstrained. Due to non-abelian interactions, the entire splitting history in the left hemisphere must be tracked to describe the distribution of the observable measured in the right hemisphere. 
}
\label{fig:GL_vs_NGL}
\end{figure}

The behavior of NGLs is distinct from that of familiar global logarithms, which we review to emphasize essential differences.  As illustrated in \Fig{fig:GL_picture}, for an observable like thrust $\tau$ \cite{Farhi:1977sg} in $e^+e^-$ collisions, enforcing $\tau \ll 1$ restricts radiation throughout the entire event.  Large logarithms of $\tau$ are generated at every order in $\alpha_s$ because of an incomplete cancellation between virtual contributions (which contribute throughout phase space) and real contributions (whose emissions are constrained by the value of $\tau$).  When contributions from all-orders in $\alpha_s$ are summed, these large infrared logarithms arrange themselves into a Sudakov form factor \cite{Sudakov:1954sw} which exponentially suppresses the small-$\tau$ region.  At double logarithmic accuracy, this Sudakov form factor $\Delta(\tau)$ is
\begin{equation}
\Delta(\tau)=e^{-\frac{\alpha_s}{\pi}C_F \log^2\tau}\,.
\end{equation}
As is well known, the Sudakov form factor expresses precisely the fact that there is no dynamics between the scale of the hard scattering, and the scale of the measurement $\tau$.

NGLs, on the other hand, are explicitly associated with dynamics occuring between two distinct scales, enforced by two distinct measurements in different regions of phase space.
Such a configuration is illustrated in \Fig{fig:NGL_picture}: here, we separate the hadronic final state of an $e^+e^-$ collision event into hemispheres, and measure the mass of the two hemispheres, $m_H$ and $m_L$.  We assume that the heavy hemisphere mass $m_H$ is larger than the light hemisphere mass $m_L$.  By requiring $m_L\ll Q$, the scattering energy, this restricts the real radiation in the right hemisphere.  However, if we consider the hierarchical case, $m_H \gg m_L$, then emissions in the left hemisphere, which occur between the scales $m_H$ and $m_L$, and are therefore unrestricted by any measurement, can contribute to the final value of $m_H$ due to non-abelian interactions. In particular, such configurations gives rise to large logarithms of the ratio of the two scales, $\log(m_H/m_L)$ in the perturbative expansion. The resummation of these large logarithms requires tracking all emissions in the left hemisphere, and therefore the NGLs do not organize themselves into a Sudakov.

A variety of approaches exist in the literature for the resummation of NGLs, each of  which requires in some form, the tracking of an infinite number of emissions to get the complete leading logarithmic series.
This was originally formulated in the leading logarithmic Monte Carlo of Dasgupta and Salam \cite{Dasgupta:2001sh}, and the Banfi-Marchesini-Smye (BMS) evolution equation \cite{Banfi:2002hw}. More recently it has been studied in the context of effective field theories and factorization from a number of different perspectives, namely the color density matrix \cite{Caron-Huot:2015bja}, the dressed gluon expansion \cite{Larkoski:2015zka}, and the SCET based approach of \cite{Becher:2015hka,Becher:2016mmh}. 

For the particular case of hemisphere masses, the leading-order BMS equation is given by
\begin{align}\label{eq:BMS_eqn_large_N}
\partial_{L}g_{ab}&=\int_\text{heavy}\frac{d\Omega_j}{4\pi}W_{ab}(j)\left(U_{abj}(L)g_{aj}g_{jb}-g_{ab}\right)\,,
\end{align}
which is an integro-differential equation for the purely non-global contribution to the cumulative cross section, $g_{ab}$, for a fundamental dipole along the directions $a,b$. Here, the angular integral for the emission $j$ is over the heavy hemisphere (out-of-jet region).  The factor
\begin{align}
W_{ab}(j)=\frac{1-\cos(\theta_{ab})}{[1-\cos(\theta_{aj})][1-\cos(\theta_{jb})]}\,,
\end{align}
which is the eikonal emission factor from the dipole with legs along the directions $a,b$, and the resummation kernel is
\begin{align}\label{eq:BMS_kernel}
U_{abj}(L)&=\exp \left[L\int_\text{light}\frac{d\Omega_q}{4\pi}\left(W_{aj}(q)+W_{jb}(q)-W_{ab}(q)\right)\right]\,.
\end{align}
Here we take $L$ to be the non-global logarithm
\begin{align}
L=\frac{\alpha_s}{\pi}N_c\log \frac{m_H}{m_L}\,.
\end{align}
The initial condition for the BMS equation is $g_{ab}(0)=1, \forall a,b$.
An analytic solution of the BMS equation is not known, however, it can be solved numerically to study its physical features, or expanded perturbatively in $\alpha_s$.

\begin{figure}
\begin{center}
\subfloat[]{\label{fig:buffer_a}
\includegraphics[width=8.75cm]{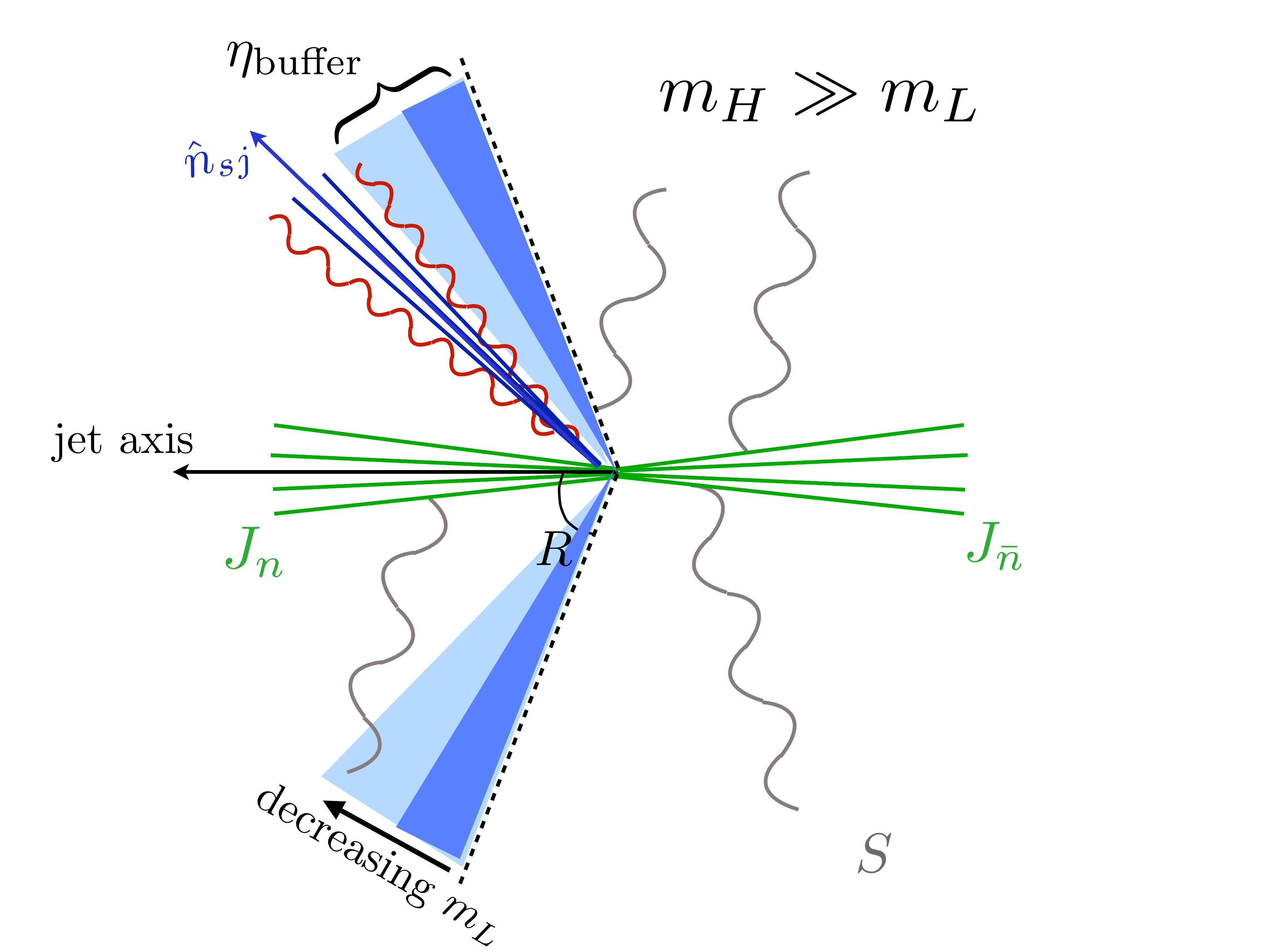}
}
\subfloat[]{\label{fig:buffer_b}
\includegraphics[width=7.5cm]{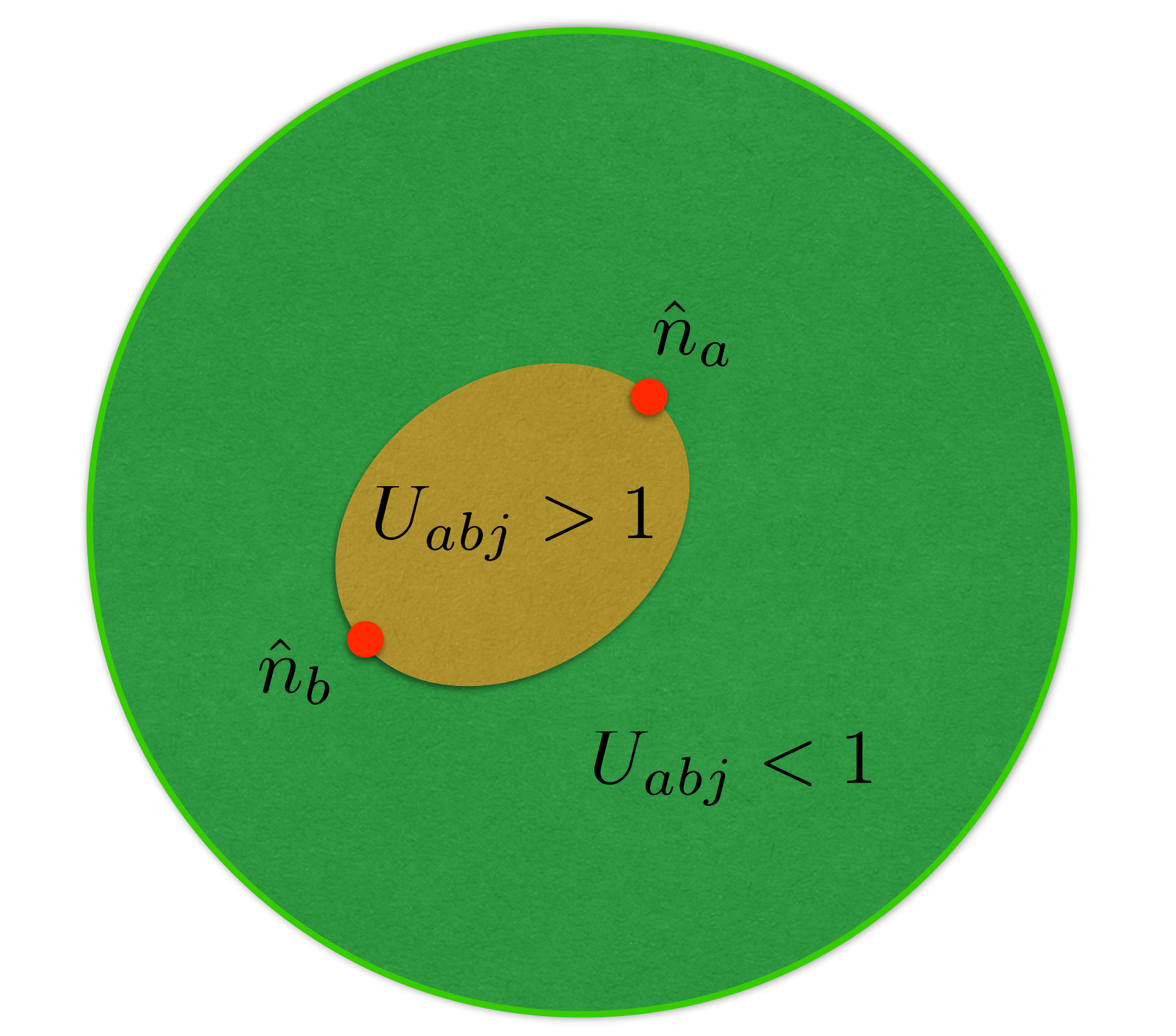}
}
\end{center}
\caption{A schematic depiction of the buffer region, which causes the probability for a dipole to emit to vanish as it approaches the edge of the jet region. The buffer region is captured in the resummed calculation, but not at fixed orders in perturbation theory. 
}
\label{fig:buffer}
\end{figure}

An important feature of NGLs which was first identified in early numerical studies, is the behavior of emissions at the boundary between the in-jet and out-of-jet regions, the so called buffer region of \Ref{Dasgupta:2002bw}. We will review in some detail this behavior, as it will also play an important role in the convergence properties of different approximations to the NGL series. If the fat jet region is approximately circular with radius $R$, the resummation factor in the BMS equation has the generic form
\begin{align}\label{eq:resummation_factor_U_for_buffer}
U_{abj}(L)&= \Bigg(1-\frac{\text{tan}^2\frac{\theta_j}{2}}{\text{tan}^2\frac{R}{2}}\Bigg)^L f_{abj}(L)\,.
\end{align}
Here $f_{abj}$ is a smooth and finite function of $j$ thoughout the jet region $J$ for all $L$, and $\theta_j$ is the angle of the fat jet axis to the emission $j$.
The factor $U_{abj}$ is bounded for a fixed $L$ for all dipoles
\begin{align} 
U_{abj}(L)\leq 2^L\,.
\end{align}
 An important consequence of the form of \Eq{eq:resummation_factor_U_for_buffer}, is that it leads to the existence of the buffer region. As $\theta_j\rightarrow R$, $U_{abj}$ vanishes, and the probability to have an emission at the boundary is zero. As $L$ increases, the size of the buffer region grows, and eventually, the only regions where emissions occur are in a small neighborhood of the initiating hard partons. This is shown schematically in \Fig{fig:buffer_a}. The physical intuition behind the buffer region is also clear. Any energetic emission at the boundary could undergo a collinear splitting, emitting a parton into the out-of-jet region, increasing the out-of-jet energy scale to that inside the jet. There is a vanishing probability for such emissions to not occur.

Since we will often refer to the buffer region in this paper, we provide a precise definition. Many definitions are of course possible, however, we prefer a definition based on the $U_{abj}$ factor, capturing the properties of the resummation factor $U$ that will feature prominently in the discussion of the fixed order series, and that always includes the $a$ and $b$ legs. We therefore define the asymptotic buffer region, $B$ as
\begin{align}\label{eq:buffer_definition}
B&=\{j\in J: U_{abj}(L)< 1\} \,, \nn\\
\bar{B}&=\{j\in J: U_{abj}(L)\geq 1\}\,.
\end{align}
We will refer to the complement of the buffer region, $\bar B$ as the active jet region, since it is the region where asymptotically one can expect emissions to populate. That the region $B$ exists is guaranteed by the form of the resummation factor in \Eq{eq:resummation_factor_U_for_buffer}. An example showing the active and buffer regions for a schematic dipole configuration is given in \Fig{fig:buffer_b}, which represents a stereographic projection of \Fig{fig:buffer_a} to better show the active and buffer regions. Here, the orange region between the two dipole legs represents the active region, which is populated by emissions, while the dark green region represents the buffer region.

However, we can also define the buffer region by any slice of the active jet region, $B_{\delta}(L)$, defined by demanding that $U$ remains less than some specified value within this slice.\footnote{This is more in keeping with the buffer region analysis of \Ref{Larkoski:2015zka}, which defined it as the full-width at half-maximum of the $U$ factor. However, since there can be regions where $U$ is exponentially growing with $L$, this definition can fail to capture the initial hard legs in an arbitrary geometry, where collinear emissions can always populate.} This region will eventually grow to the size of $B$ as $L\rightarrow\infty$. Mathematically, we have the statement that if we defined
\begin{align}
B_{\delta}(L)&=\{j\in J: U_{abj}(L)\leq 1+\delta,0<\delta<1\}\,,
\end{align}
then we have
\begin{align}
B_{\delta}(L)&\rightarrow B\text{ as  }L\rightarrow\infty\,.
\end{align}
The definition of $B$ in \Eq{eq:buffer_definition} therefore provides a natural definition for the buffer region.

As is well known, ignoring the jet integration regions, the kernel of the BMS equation including global logarithms in a conformal theory is equivalent \cite{Hatta:2008st} to the Balitsky-Kovchegov (BK) equation \cite{Balitsky:1995ub,Kovchegov:1999yj} describing unitarization and saturation effects in forward scattering. However, the presence of the jet boundary in the BMS case, which leads to the buffer region and its associated phenomena, gives rise to significantly different behavior for the solution as compared with the case of forward scattering.

\subsection{The Dressed Gluon Expansion}\label{sec:dressed_gluon_review}

In \Sec{sec:physics_BMS}, we found that for non-global observables, one must in general track the complete splitting history in the unobserved region of phase space, as was illustrated in \Fig{fig:GL_vs_NGL}. One can track this splitting history at the level of the individual partons themselves, truncating the splitting history at a fixed number of partons. This is simply the fixed-order expansion of the NGL series. An interesting question is therefore whether other expansions can be formulated.
In \Ref{Larkoski:2015zka} it was argued that one should organize this expansion not in terms of individual \emph{partonic} emissions, but in terms of identified subjets, referred to as dressed gluons, in the unobserved region of phase space. Associated with each of the identified subjets is an infinite number of unresolved gluons, captured by resummation. The resummation factor is equivalent to the $U_{abj}$ factor found in the leading-logarithmic BMS equation, and dresses the parton initiating the jet.

\begin{figure}
\begin{center}
\subfloat[]{\label{fig:twosoft}
\includegraphics[scale = .2225]{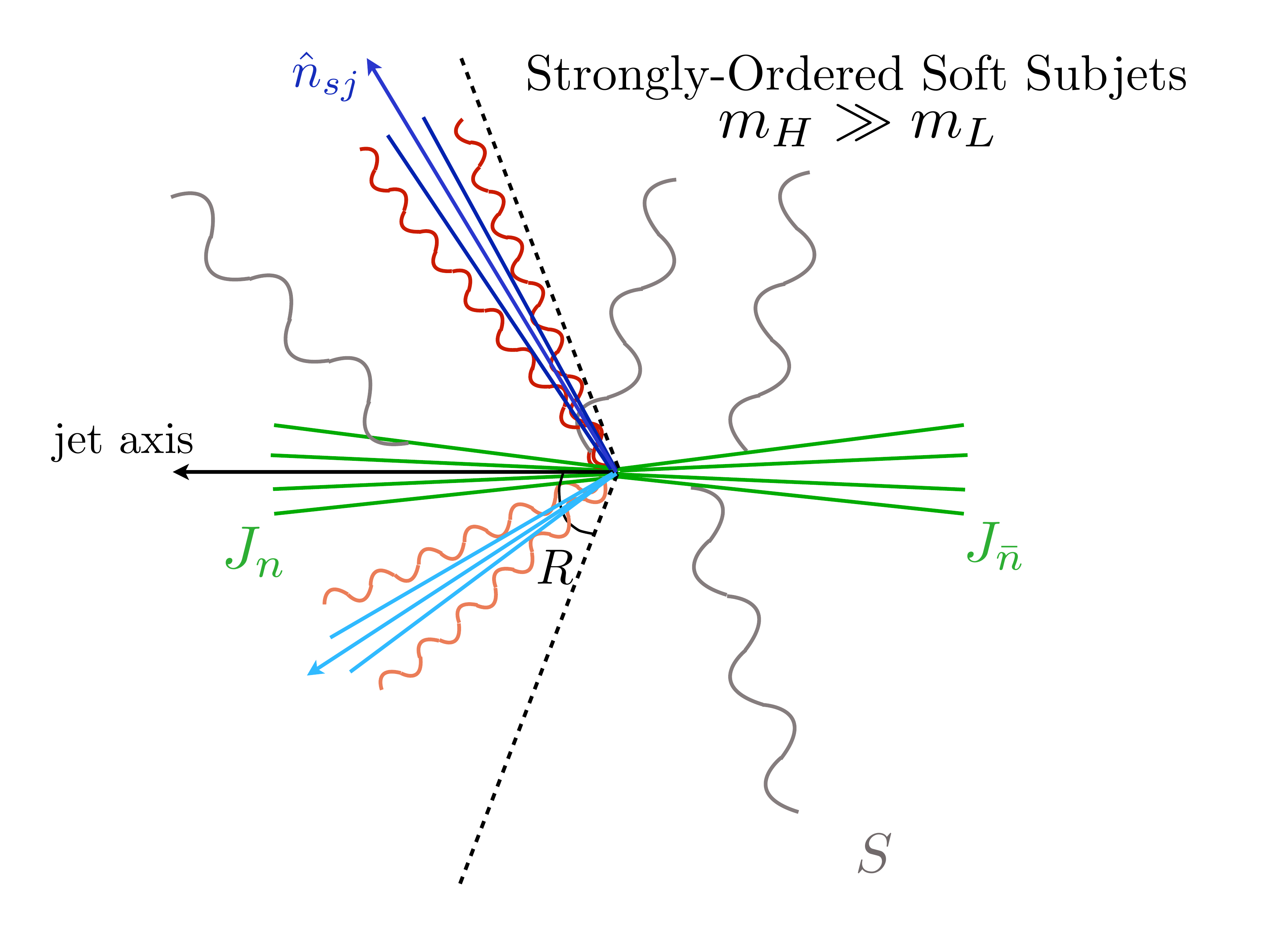}
}
$\qquad$
\subfloat[]{\label{fig:twosoft_scales}
\includegraphics[scale = 0.225]{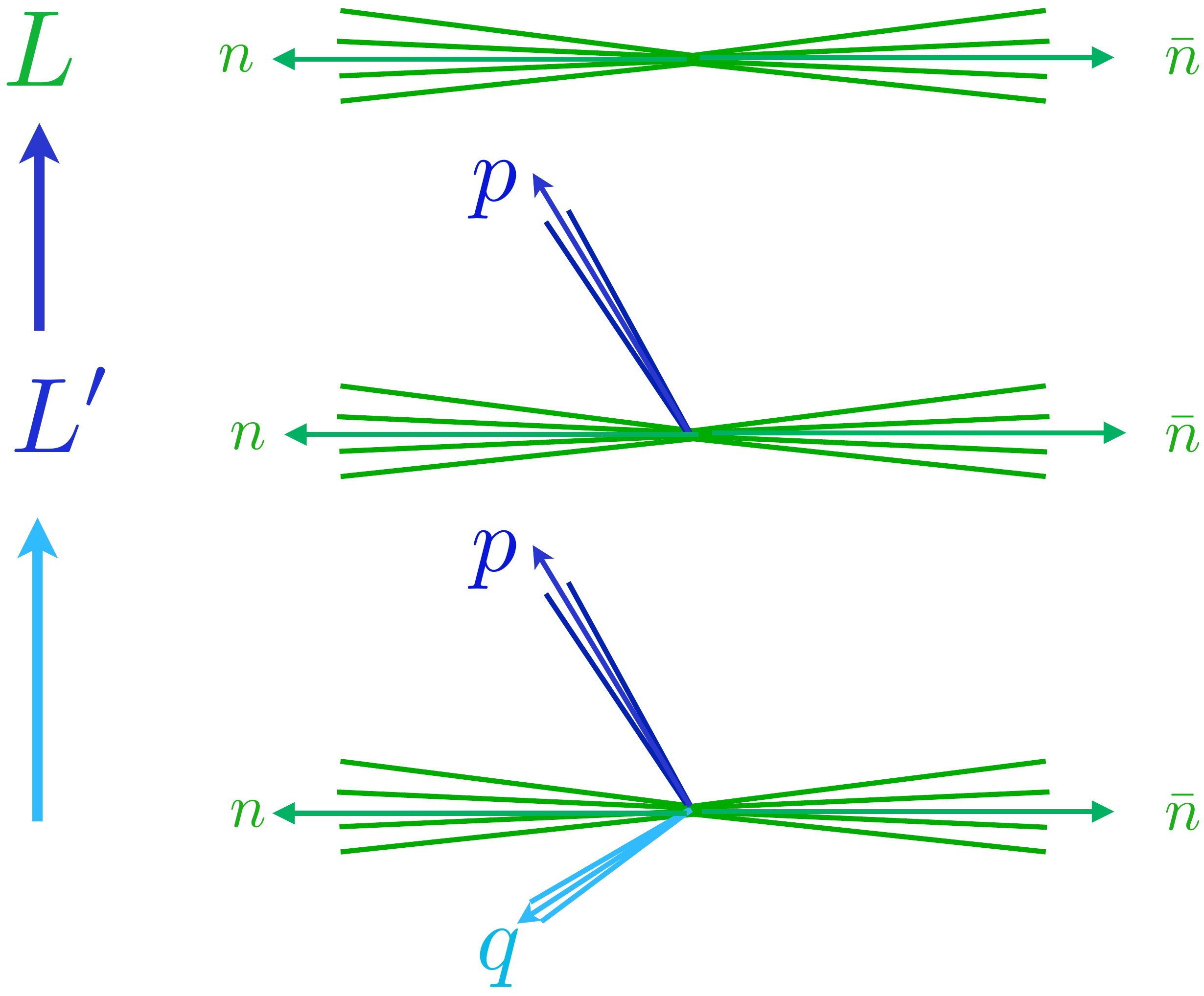}
}
\end{center}
\caption{ (a) Schematic depiction of the region of phase space defined by two strongly-ordered soft subjets, which gives rise to the leading-logarithmic two-dressed gluon expansion. (b) Illustration of the resolved subjets as a function of the resolution scale, as implemented by the matching procedure in this region of phase space. 
}
\label{fig:doublesoft}
\end{figure}

The dressed gluon expansion is therefore naturally an expansion about $g_{ab}=1$, or $L=0$, using a effective \emph{jet} state instead of a partonic state, being distinguished by the resummation. To describe the distribution at higher and higher values of $L$, more and more dressed gluons must be included in the expansion, as shown schematically in \Fig{fig:doublesoft}. The full NGL distribution is then given as a sum over all possible numbers of soft subjets, with collinear overlap regions removed. In \Ref{Larkoski:2015zka}, the dressed gluon factorization theorem was derived, and then it was hypothesized summing over multiple dressed gluons could give an expansion of the BMS equation with nice convergence properties. It was not, however, rigorously shown how the dressed gluons should be combined, nor that this was a valid expansion that converged. The goal of this paper is to clarify this expansion, and discuss some of its implications.

\begin{figure}
\begin{center}
\includegraphics[scale = .22]{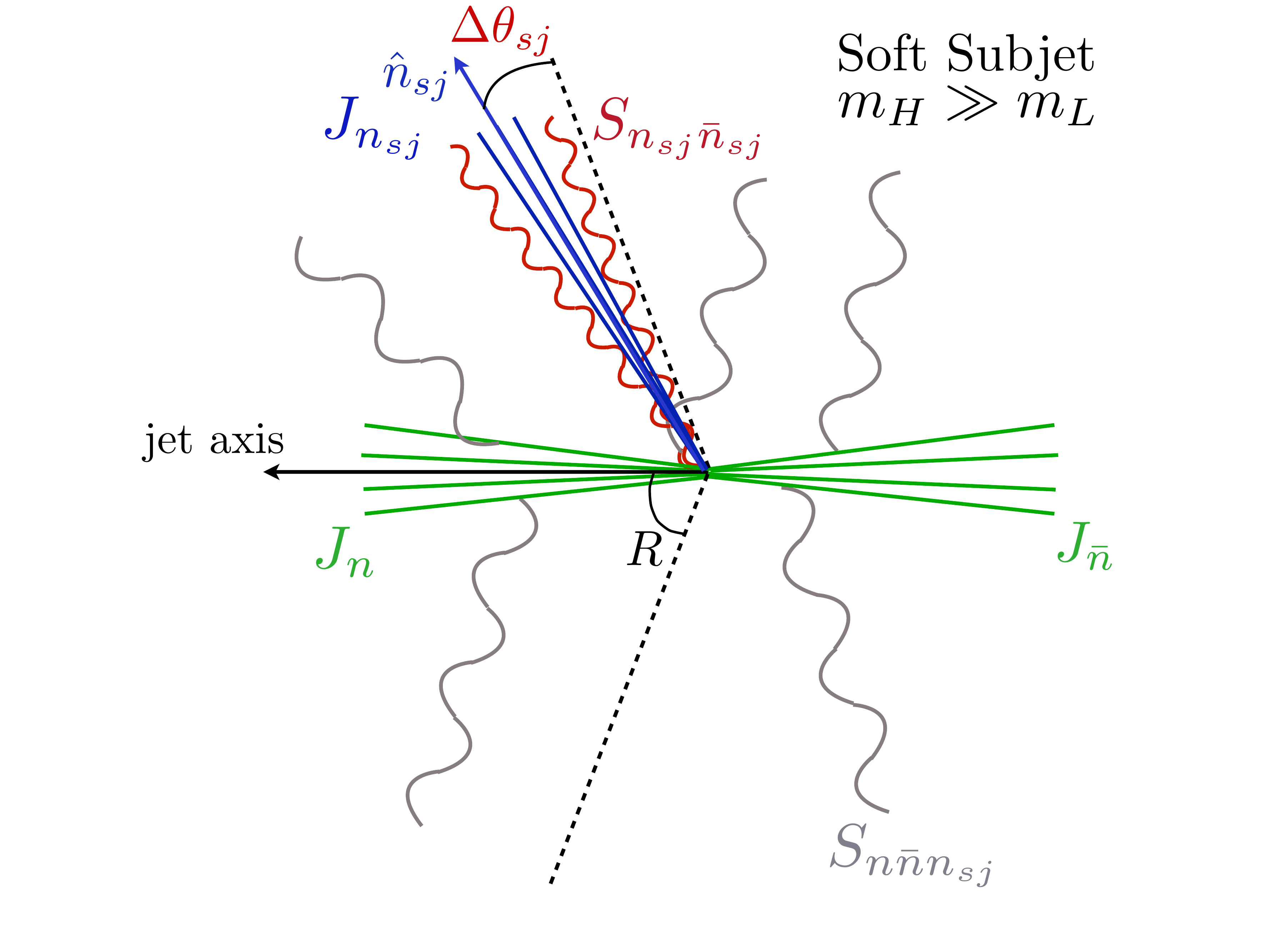}
\end{center}
\caption{A schematic depiction of the factorization theorem for a single dressed gluon, highlighting the different modes discussed in the text. Of particular importance for the discussion of NGLs, are the boundary soft modes, shown in red, which resolve the angle between the soft subjet axis, and the boundary of the jet, and resum large logarithms of this angle. 
}
\label{fig:boundary_soft_fig}
\end{figure}

Although we will not describe in detail the structure of factorization theorem for the soft subjet configuration or the construction of the dressed gluon, here we wish to emphasize several features of the factorization and associated resummation.  
For a detailed discussion, see \Ref{Larkoski:2015zka}. The schematic form of the factorization formula for a single soft subjet is given by
\begin{align}\label{eq:fact_intro}
d\sigma&=H\cdot H^{sj}_{n\bar{n}}\cdot J_{n}\otimes J_{\bar{n}}\otimes S_{n\bar{n}\sja }\otimes J_{\sja}\otimes S_{\sja\sjabar}\,,
\end{align}
which is illustrated in \Fig{fig:boundary_soft_fig}. Here we have suppressed all dependence on the resolution variables. Each of the functions appearing in \Eq{eq:fact_intro} is associated with its own resummation, which resums large logarithms of a particular scale. Of particular importance for the discussion of this paper are the \emph{boundary soft} modes, shown in red in \Fig{fig:boundary_soft_fig}, which resolve the angle between the soft jet axis, and the boundary of the jet, which we will denote $\Delta \theta_{sj}$. Large logarithms of the angle appear in the perturbative calculation, which are resummed by the boundary soft function, and are incorporated into the emission factor for the dressed gluon. As was discussed in \Ref{Larkoski:2015zka}, this gives rise to an analytic realization of the buffer region, and as will be discussed in \Secs{sec:dga}{sec:fo_fail}, this will play an important role in understanding the convergence of different expansions of the BMS equation.

In \Ref{Larkoski:2015zka} the dressed gluon expansion was studied using the energy correlation functions \cite{Larkoski:2013eya} as a resolution variable, in the specific context of the $D_2$ observable for jet substructure \cite{Larkoski:2014gra,Larkoski:2015kga}, giving rise to a precise form of the factorization formula of \Eq{eq:fact_intro} derived in SCET \cite{Bauer:2000ew, Bauer:2000yr, Bauer:2001ct, Bauer:2001yt}. However, the choice of resolution variable is immaterial. Indeed, while the dressed gluon expansion can be formulated in terms of factorization theorems identifying particular regions of phase space, \Ref{Larkoski:2015zka} also showed that the single  dressed gluon, when restricted to the leading logarithm at large $N_c$, could be directly obtained by expanding the BMS equation. To do this, we write the solution of the BMS equation as\footnote{In the context of forward scattering, the BK equation is traditionally linearized by writing $g_{ab}=1-\phi_{ab}$. Dropping non-linear terms, one recovers the BFKL equation \cite{Kuraev:1977fs,Balitsky:1978ic}, with the solution $\phi_{ab}$ being known as the pomeron. }
\begin{align}\label{eq:Defining_From_BMS_Dressed}
g_{ab}&=1+g^{(1)}_{ab} \,.
\end{align} 
Substituting this expression into the BMS equation, and expanding, one finds that $g^{(1)}_{ab}$ is the one-dressed gluon. To make this explicit, substituting \Eq{eq:Defining_From_BMS_Dressed} into the BMS equation, and expanding, we find
\begin{align}
\partial_{L}g_{ab}^{(1)}&=\int_\text{left}\frac{d\Omega_j}{4\pi}W_{ab}(j)\left(U_{abj}(L)-1\right)\,,
\end{align}
which is equivalent to the expression for the factorization theorem describing the region of phase space shown in \Fig{fig:boundary_soft_fig}, as was shown in \Ref{Larkoski:2015zka}. In \Sec{sec:pic}, we will generalize this procedure, by relating it to the method of successive approximations for solving differential equations, allowing us to give a precise definition of the dressed gluon expansion to all orders. Furthermore, we will be able to show that this expansion converges.

\section{Successive Approximations and Convergent Expansions}\label{sec:pic}

In this section, we will review the method of successive approximations for solving differential equations and contrast this with the usual perturbative expansion. While these two methods of expansion are identical for linear differential equations, such as those describing global logarithms, they in general differ for non-linear differential equations, such as those describing NGLs. After reviewing these different methods of expansion, we then show the relation between the dressed gluon expansion and the method  of successive approximations, and use the method of successive approximations to rigorously define the dressed gluon expansion at LL order.

\subsection{Fixed Order Expansions Versus Successive Approximations}

Infrared logarithms can formally be arbitrarily large. This implies that to address their behavior, one would like an expansion with an infinite radius of convergence.
That is, higher order terms in the expansion are smaller than lower order terms for arbitrary values of the logarithm, in a precise and well-defined way.  
There are of course subtleties with this convergence requirement in theories with a running coupling, but we can formally require the systematic expansion to have an infinite radius of convergence, perhaps in some appropriate conjugate space.

Na\"ively, the fixed-order $\alpha_s$ expansion for large global logarithms resummed to some accuracy satisfies this requirement.  For concreteness, we again consider calculating thrust $\tau$ \cite{Farhi:1977sg} in $e^+e^-\to$ hadrons to double logarithmic accuracy.  Thrust is a global observable: the measured value of thrust constrains all radiation throughout the entire, global, phase space.  The thrust distribution to this accuracy takes the form
\begin{equation}\label{eq:thrust}
\frac{1}{\sigma_0}\frac{d\sigma}{d\tau} = -2\frac{\alpha_s}{\pi}C_F \frac{\log \tau}{\tau} e^{-\frac{\alpha_s}{\pi}C_F \log^2\tau}\,.
\end{equation}
The exponential function has an infinite radius of convergence, and so to approximate this resummed cross section to any prescribed accuracy, one can expand in powers of $\alpha_s$ and terminate once the required accuracy has been reached.  Equivalently, one could systematically build up this resummed distribution by calculating the fixed-order cross section in the double-logarithmic limit at higher and higher orders.  Therefore, in this example, the fixed-order expansion is a good expansion that can be used to approximate to arbitrary accuracy the resummed distribution of \Eq{eq:thrust}.

More generally, logarithms that appear in the cross section of the measurement of global, infrared and collinear safe observables can often be resummed by solving renormalization group evolution equations.  In an appropriate conjugate space, these are ordinary, linear, homogeneous differential equations.  For the case of thrust, by Laplace transforming the cross section, for example, the renormalization group equations take the following schematic form
\begin{equation}\label{eq:rge}
\frac{d}{d\,\log\mu}F = \gamma F\,,
\end{equation}
where $F$ represents a part of the (Laplace-transformed) cross section.  The scale $\mu$ is the renormalization scale and $\gamma$ is the anomalous dimension of $F$.  The anomalous dimension can be calculated order-by-order in $\alpha_s$ and for observables like thrust can be written as
\begin{equation}
\gamma = \Gamma_\text{cusp}(\alpha_s)\log\frac{\mu}{\mu_0}+\gamma_\text{n-c}(\alpha_s)\,,
\end{equation}
where $\mu_0$ is a reference scale, $\Gamma_\text{cusp}(\alpha_s)$ is called the cusp anomalous dimension and $\gamma_\text{n-c}(\alpha_s)$ is called the non-cusp anomalous dimension.
The solution to \Eq{eq:rge} is then
\begin{align}\label{eq:rgesol}
F(\mu) &= F(\mu_0) e^{L(\mu,\mu_0)}\,,\nonumber\\
L(\mu,\mu_0)&=\int_{\mu_0}^\mu \frac{d\mu'}{\mu'}\, \Gamma_\text{cusp}\Big(\alpha_s(\mu')\Big)\log\frac{\mu'}{\mu_0}+\gamma_\text{n-c}\Big(\alpha_s(\mu')\Big)\,,
\end{align}
where $F(\mu_0)$ is a boundary condition. 

 Here we must draw several distinctions regarding what is meant by expansion. The expansion of $F$ in the running coupling logarithm $L$ has an infinite radius of convergence and is a good expansion for approximating the full, resummed solution of \Eq{eq:rgesol} to arbitrary accuracy, where the global $L$ is now analogous to the NGL defined with running coupling in \Eq{eq:ldef}.  This property again results from the linear nature of the renormalization group evolution equation, giving rise to the form of the solution as an exponential, \Eq{eq:rgesol}, which has an infinite radius of convergence. The anomalous dimension $\gamma$ is calculable order-by-order in $\alpha_s$, but depending on the field theory may or may not have a finite radius of expansion in $\alpha_s$.  Indeed, the cusp anomalous dimension is known to all orders in planar $\mathcal{N}=4$ SYM, where it is described by the BES equation \cite{Beisert:2006ez}. There, the perturbative expansion of the cusp anomalous dimension is known to have a finite radius of convergence.\footnote{For recent studies for convergence of $N=4$ at the amplitude level, see e.g.~\Refs{Dixon:2014voa,Dixon:2015iva}.} This has been argued to be generically true in planar theories \cite{thooft:1982}, up to issues regarding the use of renormalon free schemes. So in perturbation theory, $L$ has the expression:
\begin{align}
L(\mu,\mu_0)&=\frac{\Gamma_{\text{cusp}}^{(0)}}{4\pi}\Bigg(\int_{\mu_0}^{\mu}\frac{d\mu'}{\mu'}\alpha_s(\mu')\log\frac{\mu'}{\mu_0}\Bigg)+\frac{\Gamma_{\text{cusp}}^{(1)}}{(4\pi)^2}\Bigg(\int_{\mu_0}^{\mu}\frac{d\mu'}{\mu'}\alpha_s^2(\mu')\log\frac{\mu'}{\mu_0}\Bigg)+...\nonumber\\
&\qquad\qquad+\frac{\gamma_{n-c}^{(0)}}{4\pi}\Bigg(\int_{\mu_0}^{\mu}\frac{d\mu'}{\mu'}\alpha_s(\mu')\Bigg)+\frac{\gamma_{n-c}^{(1)}}{(4\pi)^2}\Bigg(\int_{\mu_0}^{\mu}\frac{d\mu'}{\mu'}\alpha_s^2(\mu')\Bigg)+...\,.
\end{align} 
To get to the strict fixed-order expansion of QCD, one typically also expands the running coupling order by order in perturbation theory, with the coupling evaluated at a fixed renormalization scale. The expansion of the running coupling has a finite radius of convergence, due to the Landau pole. However, this can be cured by simply leaving the fixed-order theory improved with the running coupling.\footnote{Throughout this paper, whenever we say ``fixed-order expansion,'' this can also be taken to mean ``fixed-order expansion improved with running coupling.''} This would then be equivalent to treating the $\Gamma^{(i)}$ and $\gamma^{(i)}$ as expansion parameters for the expression for $F$. What is important for our purposes is that this expansion, with the anomalous dimensions truncated to a finite order, but including the running coupling, has an infinite radius of convergence.

Without the explicit solution \Eq{eq:rgesol}, however, how do we know that the $\alpha_s$ expansion (even running coupling improved) is a systematic expansion of the cross section with an infinite radius of convergence? Furthermore, can we identify other methods of systematic expansion which go beyond the traditional expansion in $\alpha_s$? This question is of importance in the study of NGLs, as the BMS evolution equation is non-linear.

An alternative approach to a standard fixed order expansion in $\alpha_s$, which allows for the study of a more general class of differential equations is the method of successive approximations,  often also called Picard iteration \cite{CorduneanuDiff}.\footnote{In the context of integral equations, particularly of the Fredholm type, such successive approximations also go under the name of Neumann series.} Using \Eq{eq:rge} as an example, we will construct an approximate solution $F^{(n)}$, where
\begin{equation}\label{eq:picard}
F^{(n)} = \sum_{i=0}^n f^{(i)}\,,
\end{equation}
and iteratively insert it into the renormalization group equation, \Eq{eq:rge}.  That is, the equation for $F^{(n+1)}$ is
\begin{equation}\label{eq:picit}
\frac{d}{d\log\mu}F^{(n+1)} = \gamma F^{(n)}\,,
\end{equation} 
or, in terms of the $f^{(n)}$,
\begin{equation}
\frac{d}{d\log\mu}f^{(n+1)} = \gamma f^{(n)}\,.
\end{equation} 
The boundary condition sets the value of $F^{(0)}$.  Then, one looks for stationary solutions of the iterative differential equation; namely,
\begin{equation}
\lim_{n\to\infty} F^{(n)} = F\,.
\end{equation}

This limit is guaranteed to exist and is the unique solution if the differential equation satisfies a Lipschitz condition. Practically, this states that variations of the derivatives cannot be too large on any given interval.  For an arbitrary first-order, ordinary differential equation, which we can write as
\begin{equation}
\frac{dy}{dx} = g(y,x)\,,
\end{equation}
where $g$ is some function of the solution $y$ and the independent variable $x$, the Lipschitz condition is a constraint on the function $g$.  With a metric $||\cdot ||$ defined on the space of continuous functions $\{g(y,x)\}$, a Lipschitz condition is
\begin{equation} \label{eq:lip_ex}
||g(y_1,x)-g(y_2,x)|| < K |y_1-y_2|\,,
\end{equation}
where $K$ is a constant.  If this is satisfied, then the function $g$ is called Lipschitz continuous, $K$ is the Lipschitz constant, and the Picard iteration is guaranteed to converge to the unique solution in a neighborhood of $x$.  This result is known as the Picard-Lindel\"of theorem.

In the renormalization group evolution example, $g = \gamma F$, and so we can just take the metric $||\cdot ||$ to be the absolute value.  Then, the Lipschitz condition is
\begin{equation}
|\gamma| | F_1-F_2| < K |F_1-F_2|\,,
\end{equation}
which is satisfied for any $K > |\gamma|$.  Therefore, for all scales $\mu$ for which $\gamma$ is finite the Picard iteration \Eq{eq:picard} converges to the unique solution of \Eq{eq:rge}.  In particular, by solving \Eq{eq:picit}, the approximation $F^{(n)}$ is
\begin{equation}
F^{(n)} = F(\mu_0)\sum_{i=0}^n \frac{1}{n!} \left(
\int_{\mu_0}^\mu \frac{d\mu'}{\mu'} \gamma
\right)^n\,.
\end{equation}
As $n\to\infty$, this converges to the solution of \Eq{eq:rgesol}, regardless of the size of the integrated anomalous dimension.  This Picard iterated solution simply corresponds to the fixed order expansion, up to the determination of the anomalous dimensions themselves.

Generically, however, it is not true that Picard iteration corresponds to the Taylor expansion/$\alpha_s$ expansion of the solution of ordinary differential equations. In particular, this is not true for non-linear differential equations, or equations with explicit non-polynomial dependence on the independent variable, where the method of successive approximations provides a more versatile approach for identifying convergent expansions. For such non-linear equations, one expects the fixed order expansion only to have a finite radius of convergence, whereas the method of successive approximations often provides an expansion with an infinite radius of convergence, by naturally incorporating branch cuts or other singularities in its expansion.\footnote{As an explicit example which clearly demonstrates this, and shares some features with the behavior of the BMS equation, we invite the reader to compare the Taylor series and Picard iteration for the differential equation $y' = y^2 / (1+x) - y$ with $y(0)=1$.}

\subsection{The Dressed Gluon Expansion as Successive Approximations}\label{sec:DG_as_sa}

For resummation of global logarithms, there is no distinction between the perturbative expansion and the method of successive approximations, due to the linear nature of the evolution equation.  However, this is no longer true for NGLs, where the evolution equation is non-linear. 
We can identify the dressed gluon expansion, as defined in \Eq{eq:Defining_From_BMS_Dressed}, as the terms in a successive approximation about $g_{ab}=1$. In particular, the one-dressed gluon, $g_{ab}^{(1)}$ which can be computed from a factorization theorem for a single resolved jet, is exactly equivalent to the first Picard iteration of the BMS equation. This provides an interesting mathematical interpretation of the nature of the dressed gluon expansion: to describe a complicated branching history, such as that shown in the left of \Fig{fig:NGL_picture}, we can successively approximate it with subjets of increasing resolution, namely the dressed gluons. We find this correspondence between physical factorization theorems describing the number of subjets, and the method of successive approximations for solving differential equations to be quite remarkable.  Moreover, if such an expansion converges, then one can capture the physical intuition that for a fixed hierarchy of scales, an arbitrary number of emissions is unnecessary for an accurate description of the distribution. 

Using this equivalence between the first Picard iteration and the one-dressed gluon, we can now use the method of successive approximations to the BMS equation to rigorously define the dressed gluon expansion. We can define the LL dressed gluon expansion as the Picard iteration starting from $g_{ab}=1$ of the BMS equation. In particular, we define the dressed gluon expansion as
\begin{align}
g_{ab}=1+g^{(1)}_{ab}+g^{(2)}_{ab}+\cdots\,,
\end{align}
\begin{align}\label{eq:DG_recursive}
\partial_{L}g_{ab}^{(n+1)}&=\int_\text{left}\frac{d\Omega_j}{4\pi}W_{ab}(j)\left[U_{abj}(L)\left(g_{aj}^{(n)}g_{jb}^{(n)}+g_{aj}^{(n)}\sum_{i=0}^{n-1}g_{jb}^{(i)}+g_{jb}^{(n)}\sum_{i=0}^{n-1}g_{aj}^{(i)}\right)-g_{ab}^{(n)}\right]\,,
\end{align}
where $g_{ab}^{(0)}=1$ for any $a,b$.
 We therefore see that the dressed gluon can be interpreted as the kernel of the BMS equation. Here the single dressed gluon acts as the building block, and the iteration of the BMS equation defines the build up of additional dressed gluons which approximate the arbitrarily complicated gluon state. It is then natural to hypothesize that the factorization theorem for two soft subjets could act as a kernel for the NLO BMS equation. We leave a study of this to future work.

Since the dressed gluon expansion is a form of successive approximations to a non-linear differential equation, it is not an expansion in either $\alpha_s$, or $\alpha_s \log$. If we expand the dressed gluon expansion in terms of fixed order perturbation theory, it will reproduce the fixed order series.  However, it will do this in a highly non-trivial manner:  the $n$-th dressed gluon will exactly reproduce at fixed order all the NGLs up to $n$-loops, but will also contain contributions to higher loops. Given that the dressed gluon expansion reorganizes the fixed order expansion in terms of successive approximations, we can clarify the work in \Ref{Larkoski:2015zka}, where an expansion parameter was not identified. In a typical pertubative expansion, the expansion parameter can immediately be identified either as $\alpha_s$, or $\alpha_s \log$ in a resummed calculation. However, in the method of successive approximations, one can consider the expansion parameter as effectively the Lipschitz constant, $K$, which bounds the derivatives, as was discussed around \Eq{eq:lip_ex}. This is a ``worst case'' expansion parameter, giving strict upper limit on the size of the next term in the expansion, though the true size can be much smaller. Indeed, in this case the expansion parameter will be derived from bounding the kernel of the BMS equation, similar to the Lipschitz constant for ordinary differential equations.

The dressed gluon expansion of \Eq{eq:DG_recursive} is slightly different than that introduced in \Ref{Larkoski:2015zka}.  There, the dressed gluon expansion was defined similarly, as
\begin{align}
g_{ab}=1+g^{(1)}_{ab}+g^{(2)}_{ab}+\cdots\,,
\end{align}
but with the recursion 
\begin{align}\label{eq:DG_recursiveOLD}
\partial_{L}g_{ab}^{(n+1)}&=\int_\text{left}\frac{d\Omega_j}{4\pi}W_{ab}(j)\left[U_{abj}(L)\left(\sum_{i=0}^{n}g_{aj}^{(i)}g_{jb}^{(n-i)}\right)-g_{ab}^{(n)}\right]\,.
\end{align}
Each term on the right side of this expression is homogeneously the $n$-dressed gluon, with one overall angular integral.  While this form of the recursion was motivated by a possible description by increasingly differential factorization theorems, this was not explicitly proved in \Ref{Larkoski:2015zka}.  Additionally, the 1-dressed gluon of both \Eqs{eq:DG_recursive}{eq:DG_recursiveOLD} are identical, and so to distinguish them requires calculating 2- and higher dressed gluons.  Because the expansion of \Eq{eq:DG_recursive} can be directly related to Picard iteration, it will be the central focus of the remainder of the paper.  The expansion of \Eq{eq:DG_recursiveOLD}, while similar, is more challenging to formulate a proof of its convergence, and so we leave this to future work.

The fact that the dressed gluon expansion can be recast in the language of successive approximations of the BMS equation is suggestive that it is in fact a convergent expansion. Indeed, for general classes of both integral and differential equations, the method of successive approximations is known to converge. Since the BMS equation is an integro-differential equation, we will need to generalize slightly the well known convergence proofs. 
In the following sections, we will show that the BMS equation does satisfy a bounding condition, and so appropriately constructed successive approximations will converge to the unique, exact solution. We will prove that the dressed gluon expansion of \Eq{eq:DG_recursive} is such a successive approximation scheme, and that it converges. The technical details of this proof are given in \Sec{sec:dga}.

\section{Convergence of the Dressed Gluon Expansion}\label{sec:dga}

In this section we prove the main result of this paper, namely that the dressed gluon expansion of the BMS equation converges, and has an infinite radius of convergence. This section is primarily of a technical nature, and therefore readers not interested in the details of the proof can skip to the next section for applications of the dressed gluon in understanding the analytic structure of NGLs. As discussed in \Sec{sec:DG_as_sa}, from the point of view of differential equations, the dressed gluon expansion is nothing other than a rearrangement of the method of successive approximations, a standard technique for solving nonlinear differential equations. Such successive approximation techniques are known to have good convergence properties, and therefore, from this perspective, the convergence of the dressed gluon expansion is not surprising. That is, the finite truncation of the successive approximations describes the solution with a fixed accuracy within a given interval, and higher order terms will give negligible contribution, regardless of how many are dropped. Indeed, the next term in the approximation always gives an accurate assessment of the error of truncating the rest of the terms. This is opposed to an asymptotic series, where eventually higher order terms will swamp the lower order terms, and the series must be truncated at finite order, or resummed by other means.\footnote{This is not to disparage asymptotic series. These series satisfy a distinct reasonable goal: to \emph{quickly} approximate a function in some region at the cost of an upper bound of the achieveable absolute accuracy.}

The strategy of the proof is as follows. We introduce the space of mathematical functions that the BMS kernel acts upon and maps to. Then we introduce a metric on this space, and with this metric, derive a bounding condition for a collinearly regulated BMS kernel, allowing us to define a Lipschitz constant. Using this, we are then able to follow the standard logic for proving the convergence of successive approximations for a bounded kernel, which allows us to show the series is absolutely convergent on an arbitrarily sized interval $[0,L_{f}]$. Then we prove that the removal of the collinear regulator poses no difficulties for the expansion, since any solution for the BMS equation has a strict lower bound that we derive.

\subsection{The Space of Dipole Functions}\label{sec:dipole_func}

To begin, we introduce an alternative form of the BMS equation, by writing $g_{ab}=1+\phi_{ab}$, so that $\phi_{ab}$ describes the departure from $g_{ab}=1$. We then have
\begin{align}\label{eq:1_DG}
\partial_{L}\phi_{ab}&=d_{ab}(L)+\int_{J}\frac{d\Omega_j}{4\pi}W_{ab}(j)\Bigg(U_{abj}(L)\Big\{\phi_{aj}+\phi_{jb}+\phi_{aj}\phi_{jb}\Big\}-\phi_{ab}\Bigg)\,,
\end{align}
where
\begin{align}
d_{ab}(L)&=\int_{J}\frac{d\Omega_j}{4\pi}W_{ab}(j)\Big(U_{abj}(L)-1\Big)\,.
\end{align}
This form is convienent since the differential equation is now operating on what we term as a \emph{dipole} function. Loosely speaking, these are functions that map two null directions to real numbers such that they vanish when these directions become collinear. Written in this form, the BMS kernel maps dipole functions back to dipole functions. In what follows, we will restrict ourselves to a discussion of the BMS equation when the initial legs of the dipole are both inside the jet region, and will restrict our notation to this case. The proof with one leg inside, and one leg outside of the jet region proceeds analogously, but the simultaneous treatment is cumbersome. It is therefore left as an exercise for the reader.

We define the space of dipole functions in a jet region $J\subset S^2$
{\small\begin{align}
\mathcal{D}&=\{\phi_{ab}: \phi_{ab}\text{ continuously maps } J\otimes J\rightarrow \mathbb{R}; \phi_{ab}=(a\cdot b)f(a,b),  |f(a,b)|<\infty, \,\forall \,a,b \in J\} \,.
\end{align}}
\hspace{-.27cm} These dipole functions always vanish in the collinear limit ($a\cdot b\rightarrow 0$) at least as fast as $a\cdot b$. This guarantees the collinear divergences cancel in the BMS equations. It is easy to show that the space of dipole functions is also a vector space. Moreover, the function in \Eq{eq:1_DG} is a member of this dipole function space. In order to derive the convergence of the successive approximations of the BMS equation, we must have a norm on this space to judge how close two functions are. So we use essentially the standard uniform or supremum metric on $\mathcal{D}$, but with the collinear divergence factored into the norm itself
\begin{align}
\|\phi\|_{d}&=\text{sup}\Big\{\Big|\frac{\phi_{ab}}{a\cdot b}\Big|:(a,b)\in J\otimes J\Big\}\,.
\end{align}
It is a simple matter to check that this definition satisfies all the properties expected of a norm. This definition is possible since $\phi_{ab}$ is always guaranteed to have a maximum on the closure of $J\otimes J$. This norm will be useful for our purposes, since it explicitly factors in the collinear singularity, which dipole functions regulate naturally. 

\begin{figure}
\begin{center}
\subfloat[]{\label{fig:geometry_3d}
\includegraphics[width=6.75cm]{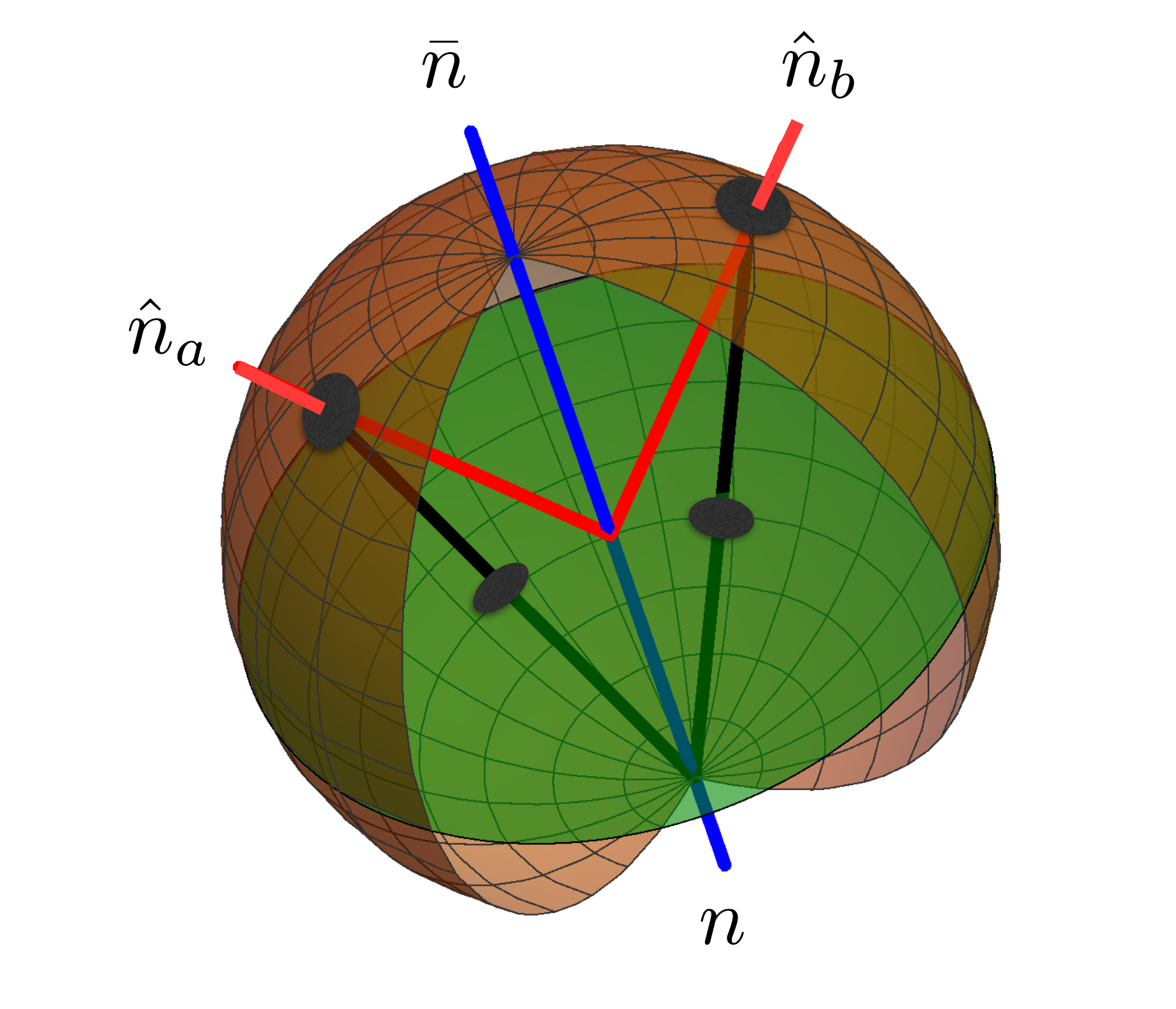}
}
\subfloat[]{\label{fig:geometry_2d}
\includegraphics[width=5.25cm]{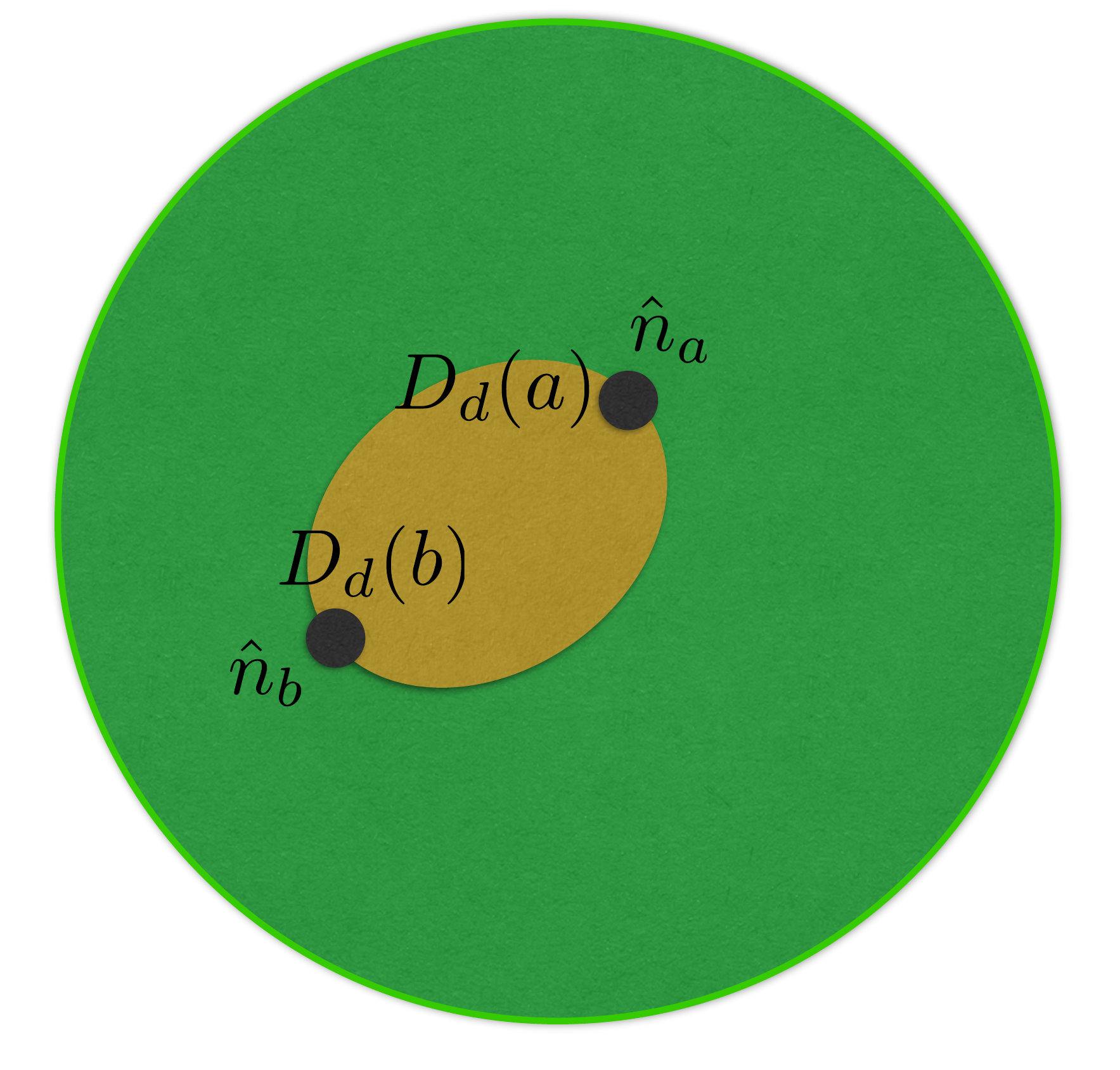}
}
\end{center}
\caption{The geometry of a jet undergoing non-global evolution. We stereographically project a circular hemisphere jet as in \Refs{Hatta:2008st,Schwartz:2014wha} onto a plane tangent to the celestial sphere at the jet axis. The green region is the heavy jet region. The orange region is the interior away from the buffer region, plotted for an example dipole given by the two dark blue points. The circles indicated the region of integration removed in the collinearly regulated BMS equation.
}
\label{fig:geometry}
\end{figure}

\subsection{The Collinearly Regulated BMS Equation}\label{sec:coll_reg_BMS}

To allow for the definition of the Lipschitz constant for the BMS equation, we will work with a collinearly regulated version of the equation, so that separate terms appearing in the equation are themselves finite. To begin, we first reorganize the form of the BMS equation. The BMS equation can be written as
\begin{align}\label{eq:BMS_eqn_Again}
\partial_L\phi_{ab}&=d_{ab}(L)+G_{ab}(L,\phi)\\
&=d_{ab}(L)+d_{ab}(L)\phi_{ab}+B_{ab}\Big(\phi;J\Big)\nonumber\\
&+\int_{J}\frac{d\Omega_j}{4\pi}W_{ab}(j)\Big(U_{abj}(L)-1\Big)\Big(\phi_{aj}+\phi_{jb}-\phi_{ab}\Big)+\int_{J}\frac{d\Omega_j}{4\pi}W_{ab}(j)U_{abj}(L)\phi_{aj}\phi_{jb}\,. \nn
\end{align}
We will call $G$ the BMS kernel. The functional $B_{ab}$ is defined as
\begin{align}
B_{ab}\Big(\phi;J\Big)&=\int_{J}\frac{d\Omega_j}{4\pi}W_{ab}(j)\Big(\phi_{aj}+\phi_{jb}-\phi_{ab}\Big)\,.
\end{align}
We write the BMS equation as above to explicitly single out the $B$-functional. When $J=S^2$, the whole celestial sphere, then $B_{ab}$ is conformally related to the position space form of the kernel of the BFKL equation; see \Refs{Mueller:1994gb, Marchesini:2003nh, Hatta:2008st, Avsar:2009yb}, and also see \Ref{Marchesini:2004ne} for the construction of solutions based on this kernel. 

The BFKL kernel has unbounded eigenvalues when acting on the space of dipole functions, rendering it impossible to directly bound $B_{ab}$. Thus, it will be important to introduce the collinearly regulated BMS kernel, where we replace the jet region $J$ by the jet region with two small regions surrounding the legs of the dipoles removed, as shown in \Fig{fig:geometry}. To discuss these configurations, we introduce the following notation. We define $D_{\delta}(p)$ to be the open disc of radius $\delta$ about point $p$.
The collinearly regulated jet region is then defined as
\begin{align}
R^{\delta}_{ab}&=J-D_{\delta}(a)\cup D_{\delta}(b)\,, \nonumber \\
\overline{R}^{\delta}_{ab}&=D_\delta(a)\cup D_\delta(b)\,.
\end{align}
This allows us to define a collinearly regulated BMS equation
\begin{align}\label{eq:BMS_kernel_col_reg}
G_{ab}^{\delta}(L,\phi)&=d_{ab}^{\delta}(L)\phi_{ab}+B_{ab}^{\delta}\Big(\phi;R^{\delta}_{ab}\Big)\\
&\hspace{-0.5cm}+\int_{R^{\delta}_{ab}}\frac{d\Omega_j}{4\pi}W_{ab}(j)\Big(U_{abj}(L)-1\Big)\Big(\phi_{aj}+\phi_{jb}-\phi_{ab}\Big)+\int_{R^{\delta}_{ab}}\frac{d\Omega_j}{4\pi}W_{ab}(j)U_{abj}(L)\phi_{aj}\phi_{jb}\,, \nn
\end{align}
where
\begin{align}
d_{ab}^{\delta}(L)&=\int_{R^{\delta}_{ab}}\frac{d\Omega_j}{4\pi}W_{ab}(j)\Big(U_{abj}(L)-1\Big)\,.
\end{align}
This collinearly regulated version of the BMS equation will be our main focus of study, and will allow us to prove convergence of the dressed gluon expansion.
Cutting off the collinear region explicitly makes the BFKL kernel's contribution to the BMS equation a bounded linear functional. Note that the BMS equation is collinear finite, however, this occurs due to a cancellation of real and virtual terms. For the collinearly regulated BMS equation, each term is separately bounded. In \Sec{sec:coll_reg}, we will argue how the collinear regulator can be removed.

\subsection{Bounding the Collinearly Regulated BMS Kernel}\label{sec:bound_kernel}

Having defined a collinearly regulated version of the BMS equation, given in \Eq{eq:BMS_kernel_col_reg}, in this section we derive a Lipschitz condition of the form
\begin{align}
\Big|G_{ab}^{\delta}(L,\phi)-G_{ab}^{\delta}(L,\psi)\Big|&\leq(a\cdot b)K^\delta \Big\|\phi-\psi\Big\|_{d}
\end{align}
for the BMS kernel.

We will work with the space of bounded dipole functions near a specified dipole function $v_{ab}$, defined as
\begin{align}
\mathcal{D}_M(v_{ab})=\{\phi_{ab}\in \mathcal{D} \quad \& \quad  \|\phi-v\|_{d}\leq M\} \,.
\end{align}
Though on the space of dipole functions, no collinear regularization is necessary, we will be able to show that the collinearly regulated BMS kernel obeys a Lipschitz condition on $\mathcal{D}_M(v_{ab})$ if $v_{ab}$ itself is bounded on the dipole jet region.

In bounding the difference of the BMS kernel acting on two dipole functions, we will make use of the following functions, which we define for notation convenience
\begin{align}
\mathcal{A}&=\text{max}_{a,b\in J}[a\cdot b]\,, \nonumber \\
\label{eq:dmax_definition}d_{max}(L)&=\text{max}_{a,b}\int_{j\in J}\frac{W_{ab}(j)}{a\cdot b}\Big|U_{abj}(L)-1\Big|\,,\nonumber\\
A(J)&=\int_J\frac{d\Omega_j}{4\pi}\,,\nonumber \\
\kappa^\delta_J&=A(J)\Big(\text{max}_{a,b\in J}\text{max}_{j\in R^{\delta}_{ab}}\frac{W_{ab}(j)}{a\cdot b}\Big)\,.
\end{align}
We note that
\begin{align}
\left|d_{ab}^{\delta}(L)\right|\leq (a\cdot b) d_{max},\qquad\forall \delta\geq 0\,,
\end{align}
and therefore $d_{max}$ is also a maximal estimate for any collinearly regulated jet regions, since by construction it has no collinear divergences on $J$. 

We now examine the difference
\begin{align} 
\label{eq:BMS_kernel_ineq_start_II}\Big|G_{ab}^{\delta}(L,\phi)-G_{ab}^{\delta}(L,\psi)\Big|\,,
\end{align}
which we would like to bound.
Using the triangle inequality, it is sufficient to derive a bound on each term in \Eq{eq:BMS_kernel_col_reg}.
Beginning with the three linear terms, and proceeding in the order that they appear in \Eq{eq:BMS_kernel_col_reg}, we have
\begin{align}
&\left|d_{ab}^{\delta}(L)\right|\phi_{ab}\leq (a\cdot b)\mathcal{A} d_{max} \|\phi\|_d\,,
\end{align}
\begin{align}
\Big|B_{ab}^{\delta}\Big(\phi;R^{\delta}_{ab}\Big)-B_{ab}^{\delta}\Big(\psi;R^{\delta}_{ab}\Big)\Big|   \leq   3(a\cdot b)\mathcal{A}\kappa^\delta_J&\|\phi-\psi\|_{d}\,,
\end{align}
\begin{align}
&\int_{R^{\delta}_{ab}}\frac{d\Omega_j}{4\pi}W_{ab}(j)\Big(U_{abj}(L)-1\Big)\Big|\phi_{aj}+\phi_{jb}-\phi_{ab}-\psi_{aj}-\psi_{jb}+\psi_{ab}\Big|\nonumber \\
&
\hspace{9cm}
\leq 3(a\cdot b)\mathcal{A}d_{max}\|\phi-\psi\|_{d}\,.
\end{align}
The final nonlinear term can be bounded as
\begin{align}
\int_{R^{\delta}_{ab}}\frac{d\Omega_j}{4\pi}&W_{ab}(j)U_{abj}(L)\Big|\phi_{aj}\phi_{jb}-\psi_{aj}\psi_{jb}\Big|\nonumber\\
&=\frac{a\cdot b}{2}\int_{R^{\delta}_{ab}}\frac{d\Omega_j}{4\pi}U_{abj}(L)\Bigg|\frac{\Big(\phi_{aj}-\psi_{aj}\Big)}{a\cdot j}\frac{\Big(\psi_{jb}+\phi_{jb}\Big)}{j\cdot b}+\frac{\Big(\phi_{aj}+\psi_{aj}\Big)}{a\cdot j}\frac{\Big(\phi_{jb}-\psi_{jb}\Big)}{j\cdot b}\Bigg|\nonumber\\
&\leq (a\cdot b) \mathcal{U}(L) \Big\|\phi+\psi\Big\|_{d}\Big\|\phi-\psi\Big\|_{d}\,,
\end{align}
where we have defined
\begin{align}
\mathcal{U}(L)&=\text{max}_{a,b\in J}\int_{J}\frac{d\Omega_j}{4\pi}U_{abj}(L)\,.
\end{align}
Note that because $U_{abj}(L)$ is positive definite, the integral over the full jet region $J$ is always larger than over the punctured jet region, $R^{\delta}_{ab}$.
Collecting the different contributions, we find
\begin{align}
\label{eq:BMS_kernel_ineq_start_III}\Big|G_{ab}^{\delta}(L,\phi)-G_{ab}^{\delta}(L,\psi)\Big|&\leq(a\cdot b)\Big(4\mathcal{A}d_{max}(L)+3\mathcal{A}\kappa^\delta_J+\mathcal{U}(L) \Big\|\phi+\psi\Big\|_{d}\Big) \Big\|\phi-\psi\Big\|_{d}\,.
\end{align}
Thus when $L$ is restricted to a fixed interval $[L_i,L_f]$, we can find the maximizing $L$, and if $\phi,\psi\in\mathcal{D}_M(v_{ab})$, we obtain the desired Lipschitz condition:
\begin{align}\label{eq:BMS_kernel_lipschitz}
\Big|G_{ab}^{\delta}(L,\phi)-G_{ab}^{\delta}(L,\psi)\Big|&\leq(a\cdot b)K^\delta \Big\|\phi-\psi\Big\|_{d}\,,
\end{align}
where the Lipschitz constant $K^\delta$ is defined as
\begin{align}
K^\delta&=4\mathcal{A}\Big(\text{max}_{L\in[L_i,L_f]}d_{max}(L)\Big)+3\mathcal{A}\kappa^\delta_J+2\Big(\text{max}_{L\in[L_i,L_f]}\mathcal{U}(L)\Big) \Big(M+\|v\|_d\Big)\,.
\end{align}

\subsection{Proof of Convergence}\label{sec:proof}

Having derived a Lipschitz constant for the collinearly regulated BMS equation, we now proceed with a proof of convergence for the dressed gluon expansion. This will be done in two steps. We first proof a local convergence of the solution, and then show that this solution can be arbitrarily continued to any given fixed interval, completing the proof of convergence for the regulated BMS equation.

\subsection*{Local Existence}
We begin by proving local existence. For clarity, we write the proof in the form of bulleted steps.
\begin{itemize}
\item Focus on $L$ in an interval $I_{L_0}^{A}=[L_0-A,L_0+A]$, and we examine the initial value problem stated as an integral equation:
\begin{align}
\phi_{ab}^{\delta}&=v_{ab}+\int_{L_0}^{L}dL'\,G_{ab}^{\delta}(L',\phi^{\delta}).
\end{align}
We seek to find the solution $\phi_{ab}^{\delta}$ within the space $\mathcal{D}_{M}(v_{ab})$ for $|L-L_0|$ sufficiently small.
\item Define the successive approximations:
\begin{align}
\phi_{0;ab}^{\delta}&=v_{ab}\,,\nonumber\\
\label{eq:successive_approx_BMS_CREG}\phi_{n;ab}^{\delta}&=v_{ab}+ \int_{L_0}^{L}dL'\,G_{ab}^{\delta}(L',\phi_{n-1}^{\delta})\,.
\end{align}
\item Let $\lambda=\text{min}\{A,\frac{M}{C}\}$, $C=\text{sup}|G_{ab}^{\delta}(L,\psi)|$, for $L\in I_{L_0}^{A}$, $a,b\in J$, and $\psi\in \mathcal{D}_{M}(v_{ab})$. From the Lipschitz condition \eqref{eq:BMS_kernel_lipschitz}, such a $C$ exists, and $C\leq \mathcal{A}K^{\delta} (M+\|v\|_d)$. We now restrict to the subinterval $I_{L_0}^{\lambda}\subset I_{L_0}^{A}$. Note that we have maximized with respect to the opening angle also with the factor of $\mathcal{A}$. 
\item The successive approximations remain within $\mathcal{D}_M(v_{ab})$. We show this by induction:
\begin{itemize} 
\item The first approximation is in $\mathcal{D}_M(v_{ab})$:\begin{align}
    |\phi_{1;ab}^{\delta}-\phi_{0;ab}^{\delta}|\leq C|L-L_0|\leq M\,.\label{eq:first_iteration_bound}
  \end{align}
  Since this is true for all $a,b\in J$, we have $\phi_{1;ab}^{\delta}\in \mathcal{D}_M(v_{ab})$ for $L\in I_{L_0}^{\lambda}$.
\item Assume $\phi_{k;ab}^{\delta}\in\mathcal{D}_M(v_{ab})$ for $k<n$. Then from \eqref{eq:successive_approx_BMS_CREG}:
  \begin{align}
    |\phi_{n;ab}^{\delta}-\phi_{0;ab}^{\delta}|\leq C|L-L_0|\leq M\,.
  \end{align} 
  So $\phi_{n;ab}^{\delta}\in\mathcal{D}_M(v_{ab})$.
\end{itemize}
\item We can show inductively that
\begin{align}\label{eq:induction_bound}
 \|\phi_{n;ab}^{\delta}-\phi_{n-1;ab}^{\delta}\|_d\leq\frac{C}{K^{\delta}}\frac{\Big(K^{\delta}|L-L_0|\Big)^{n}}{n!}\,.
\end{align}
 The proof of this proceeds as follows:
\begin{align}
|\phi_{n;ab}^{\delta}-\phi_{n-1;ab}^{\delta}|&=\Big|\int_{L_0}^{L}dL'\,G_{ab}^{\delta}(L',\phi_{n-1}^{\delta})-\int_{L_0}^{L}dL'\,G_{ab}^{\delta}(L',\phi_{n-2}^{\delta})\Big|\nonumber\\
 &\leq (a\cdot b)K^\delta \int_{L_0}^{L}dL'\Big\|\phi_{n-1}-\phi_{n-2}\Big\|_{d} 
\end{align}
Now using the induction hypothesis:
\begin{align}
\Big\|\phi_{n-1}(L')-\phi_{n-2}(L')\Big\|_{d}=\frac{C}{K^{\delta}}\frac{(K^\delta)^{n-1}|L'-L_0|^{n-1}}{(n-1)!}
\end{align}
we achieve \Eq{eq:induction_bound}. The first step in the induction was already established in \Eq{eq:first_iteration_bound}.
\item Thus by the Weierstrauss M-test, the series:
\begin{align}
\phi_{ab}^{\delta}(L)&=v_{ab}+\sum_{n=1}^{\infty}\Big(\phi_{n;ab}^{\delta}(L)-\phi_{n-1;ab}^{\delta}(L)\Big)\,,
\end{align}
converges. The $n-$th partial sum of this series is just the $\phi_{n;ab}^{\delta}$.
\item We now show $\phi_{ab}^{\delta}$ solves the BMS equation. This follows from:
\begin{align}
|G_{ab}^{\delta}(L,\phi_n^\delta)-G_{ab}^{\delta}(L,\phi^\delta)|\leq  \mathcal{A}K^{\delta} \|\phi_n^\delta-\phi^\delta\|\,.
\end{align}
This implies the application of $G_{ab}^{\delta}$ on the sequence $\{\phi_{n;ab}^{\delta}\}$ converges since the sequence itself does. Thus we can take the limit $n\rightarrow \infty$ on both sides of \eqref{eq:successive_approx_BMS_CREG}, and substitute in $\phi^{\delta}_{ab}$:
\begin{align}
\label{eq:BMS_CREG}\phi_{ab}^{\delta}&=v_{ab}+\int_{L_0}^{L}dL'\,G_{ab}^{\delta}(L',\phi^{\delta})\,.
\end{align}
\end{itemize}
With more work, one can also show uniqueness.

\subsection*{Global Existence}

Now we fix an interval $[L_i,L_f]$ and an $M$. We wish to show a solution exists on this predefined interval. We start with the initial value problem with:
\begin{align}
\phi_{ab}^{\delta}&=v_{ab}+\int_{L_i}^{L}dL'\,G_{ab}^{\delta}(L',\phi^{\delta})\,.
\end{align}
The idea of the proof is to keep continuing the local solution given above until we cover the whole interval.

\begin{itemize}
\item{{\bf{First Interval:}}}

We construct the Lipschitz constant $K_1^{\delta}$ for the set $L\in [L_i,L_f]$ and $\phi\in \mathcal{D}_{M}(v)$. We can solve the BMS equation on the interval:
\begin{align}
[L_i,L_i+\lambda_1)
\end{align}
Where:
\begin{align} 
\lambda_1&=\text{min}\Big\{L_f-L_i,\frac{M}{C_1}\Big\}\,,\\
C_1&=\text{sup}|G^{\delta}_{ab}(L,\phi)|, \qquad L\in[L_i,L_f]\text{ and } \phi\in\mathcal{D}_{M}(v)\,.
\end{align}
If $\lambda_1=L_f-L_i$, we are done. Therefore, assume otherwise. We note that:
\begin{align}
\frac{1}{ \mathcal{A}K_1^{\delta}}\leq\lambda_1\,.
\end{align}

\item{{\bf{Second Interval:}}}

We now take the limit, $L\rightarrow L_0+\lambda_1$, and given that on the first interval, the solution is bounded and a differentiable function of $L$, the limit to the endpoint exists and is continuous. Let $v^{(2)}_{ab}=\phi_{ab}^{\delta}(L_0+\lambda_1)$ and consider the new initial value problem:
\begin{align}
\phi_{ab}^{\delta}&=v_{ab}^{(2)}+\int_{L_0+\lambda_1}^{L}dL'\,G_{ab}^{\delta}(L',\phi^{\delta})\,.
\end{align}
We construct the Lipschitz constant $K_2$ for functions $\phi\in \mathcal{D}_{M}(v^{(2)})$, keeping in mind we already maximized the functions $\mathcal{U}$ and $d_{max}$ on the interval $[L_i,L_f]$. Now we can solve the BMS equation on the interval:
\begin{align}
[L_i+\lambda_1,L_i+\lambda_1+\lambda_2)
\end{align}
Where:
\begin{align} 
\lambda_2&=\text{min}\Big\{L_f-L_i-\lambda_1,\frac{M}{C_2}\Big\}\,,\\
C_2&=\text{sup}|G^{\delta}_{ab}(L,\phi)|, \qquad L\in[L_i,L_f]\text{ and } \phi\in\mathcal{D}_{M}(v^{(2)})\,.
\end{align}
If $\lambda_2=L_{max}-L_0-\lambda_1$, we are done. Therefore, assume otherwise. Now again we have:
\begin{align}
\frac{1}{\mathcal{A}K_2^{\delta}}\leq \lambda_2\,.
\end{align}
Importantly,
\begin{align} 
K_2^{\delta}-K_1^{\delta}&=2\Big(\text{max}_{L\in[L_i,L_f]}\mathcal{U}(L)\Big)(\|v^{(2)}\|_d-\|v\|_d)\leq \alpha M\,,
\end{align}
where the constant $\alpha$ is defined as
\begin{align}
\alpha&=2\Big(\text{max}_{L\in[L_i,L_f]}\mathcal{U}(L)\Big)\,.
\end{align} 
This is due to the fact that the dependence on $D_{M}(v)$ appears linearly in Eq. \eqref{eq:BMS_kernel_lipschitz}, and $v^{(2)}\in \mathcal{D}_{M}(v)$. We therefore conclude:
\begin{align}
\frac{1}{ \mathcal{A}K_1^{\delta}+\alpha \mathcal{A} M}\leq \lambda_2\,.
\end{align}

\item{{\bf{$n$-th Interval:}}}

We continue constructing new intervals. Let $v^{(n)}$ be the initial value for the $n$-th interval. Importantly, on every new interval, the new Lipschitz constant is related to the old one by the inequality:
\begin{align}
K_n^{\delta}\leq K_{n-1}^{\delta}+\alpha M\,.
\end{align}
We solve this recursion to conclude:
\begin{align}
K_n^{\delta}\leq K_{1}^{\delta}+(n-1)\alpha M\,.
\end{align}
And, unless we hit $L_{f}$, the new interval is given by:
\begin{align}
\Big[L_i+\sum_{k=1}^{n-1}\lambda_k, L_i+\sum_{k=1}^{n}\lambda_k\Big)
\end{align}
However, this sum is bounded from below by a harmonic sum:
\begin{align}
\sum_{k=1}^{n}\frac{1}{ \mathcal{A}K_{1}^{\delta}+(k-1)\alpha  \mathcal{A} M}\leq \sum_{k=1}^{n}\lambda_k
\end{align}
Such harmonic sums grow arbitrarily large. Thus we conclude after a finite number of continuations, we will eventually hit $L_{f}$, and we now have a solution on the whole interval $[L_i,L_{f}]$. This solution is also bounded, for if $N$ is the integer such that $\sum_{k=1}^{N-1}\lambda_k<L_f-L_i\leq \sum_{k=1}^{N}\lambda_k$, then the continued solution is in the set $\mathcal{D}_{N M}(v)$, since the norm of each initial value $v^{(k)}$ for the $k$-th interval is bounded as $\|v^{(k)}\|_{m}\leq M+\|v^{(k-1)}\|_m$. It is interesting to note that the dependence on the collinear regulator is entirely a feature of the first Lipschitz constant. This is essentially due to the fact that the bound on the nonlinear terms do not depend on the collinear regularization.

\end{itemize}

\subsection{Removing the Collinear Regulator}\label{sec:coll_reg}

We now indicate how one can justify removing the collinear regulator. That the limit to $\delta\rightarrow 0$ is smooth is entirely reasonable, since the regulator itself appears as a boundary of an integral, and so its explicit functional dependence in the BMS kernel is differentiable. First we will demonstrate the solutions to the BMS equation are bounded from below for all values of the collinear regulator. Then if we remove the collinear regularization anywhere the BMS solution is a decreasing function, the solutions will remain bounded, and we will have a well-defined limit. Indeed, we find that one should expect faster decay of the BMS solution for smaller values of the regulator.

To derive the lower bound, we make use of the comparison theorem for differential equations (see for instance, \Ref{CorduneanuDiff}), which states that if we have a differential inequality
\begin{align}
\dot{x}&\geq w(t,x)\,,\,x(t_0)=x_0\,,
\end{align}
then solutions to the differential equation
\begin{align}
\dot{y}&= w(t,y)\,,\,y(t_0)=x_0\,,
\end{align}
will bound from below solutions to the differential inequality with identical initial conditions. Thus if we have a differential equation
\begin{align}
\dot{x}&= f(t,x)\,.
\end{align} 
It suffices to construct an $\omega(t,x)$ that bounds from below the original $f(t,x)$ to achieve a lower bound on solutions. It is simple to show that $\phi_{ab}=-1$ is a fixed point for all values of $\delta$ in the regulated BMS equation \eqref{eq:BMS_kernel_col_reg}. Thus we return to the original form of the regulated BMS equation
\begin{align}
\label{eq:BMS_C_REG_Near_n1}\partial_Lg_{ab}^{\delta}&=\int_{R^{\delta}_{ab}}\frac{d\Omega_j}{4\pi}W_{ab}(j)\Big(U_{abj}(L)g_{aj}^{\delta}g_{jb}^{\delta}-g_{ab}^{\delta}\Big)\,.
\end{align}
Noting that
\begin{align} 
W_{ab}(j)U_{abj}(L)g_{aj}g_{jb}\geq 0\,,
\end{align}
we can drop the nonlinear term, cancel the explicit $L$ dependence, and derive the differential inequality
\begin{align}
\partial_Lg_{ab}^{\delta}&\geq-\gamma_{ab}(\delta)g_{ab}^{\delta}\,,\\
\gamma_{ab}(\delta)&=\int_{R^{\delta}_{ab}}W_{ab}(j).
\end{align}
Our truncation of the regulated BMS equation now satisfies the conditions for the comparison theorem, so the solution to the differential equation
\begin{align}
\partial_Lg_{ab}^{\delta}&=-\gamma_{ab}(\delta)g_{ab}^{\delta}\,,
\end{align}
will bound from below solutions to \eqref{eq:BMS_C_REG_Near_n1}, as long as they have the same initial condition. This then implies, if $g_{ab}^{\delta}(0)=1$,
\begin{align}\label{eq:bounding_solutions_to_BMS}
g_{ab}^{\delta}(L)\geq \exp \left[-L\gamma_{ab}(\delta)\right]\,.
\end{align}
We can compute the dependence of $\gamma_{ab}(\delta)$ on the cutoff $\delta$ as it is determined by the collinear divergences of gauge theories. When both $a,b$ are in the jet we have
\begin{align}\label{eq:bounding_solutions_to_BMS_gamma_in_in}
\gamma_{ab}(\delta)=-2\,\log\,\delta+\mathcal{O}(1),\text{ if } \theta_{ab}>\delta\,,
\end{align}
where $\theta_{ab}$ is the angle between directions $a$ and $b$. When only one leg is inside the jet we have
\begin{align}\label{eq:bounding_solutions_to_BMS_gamma_in_out} 
\gamma_{ab}(\delta)=-\log\,\delta+\mathcal{O}(1)\,.
\end{align}
As $\delta\rightarrow 0$, the logarithmic cutoff eventually dominates any dipole opening angle, and this becomes a bound on the solutions of the unregulated BMS equation:
\begin{align}
\lim_{\delta\rightarrow 0}g_{ab}^{\delta}(L)\geq 0\,,\forall L \,.
\end{align}
As long as the solution to the BMS equation has a negative derivative, we can remove the collinear regulator and ensure that the solutions to the BMS equation are bounded from below for all $\delta$ and $L$. As this is the case for physical initial conditions, we can then take $\delta\rightarrow 0$ in the above successive approximations, constructing a solution to the BMS equation via a series with an infinite radius of convergence.

\begin{figure}
\begin{center}
\subfloat[]{\label{fig:b_t_b_Cutoff_a}
\includegraphics[width=7.5cm]{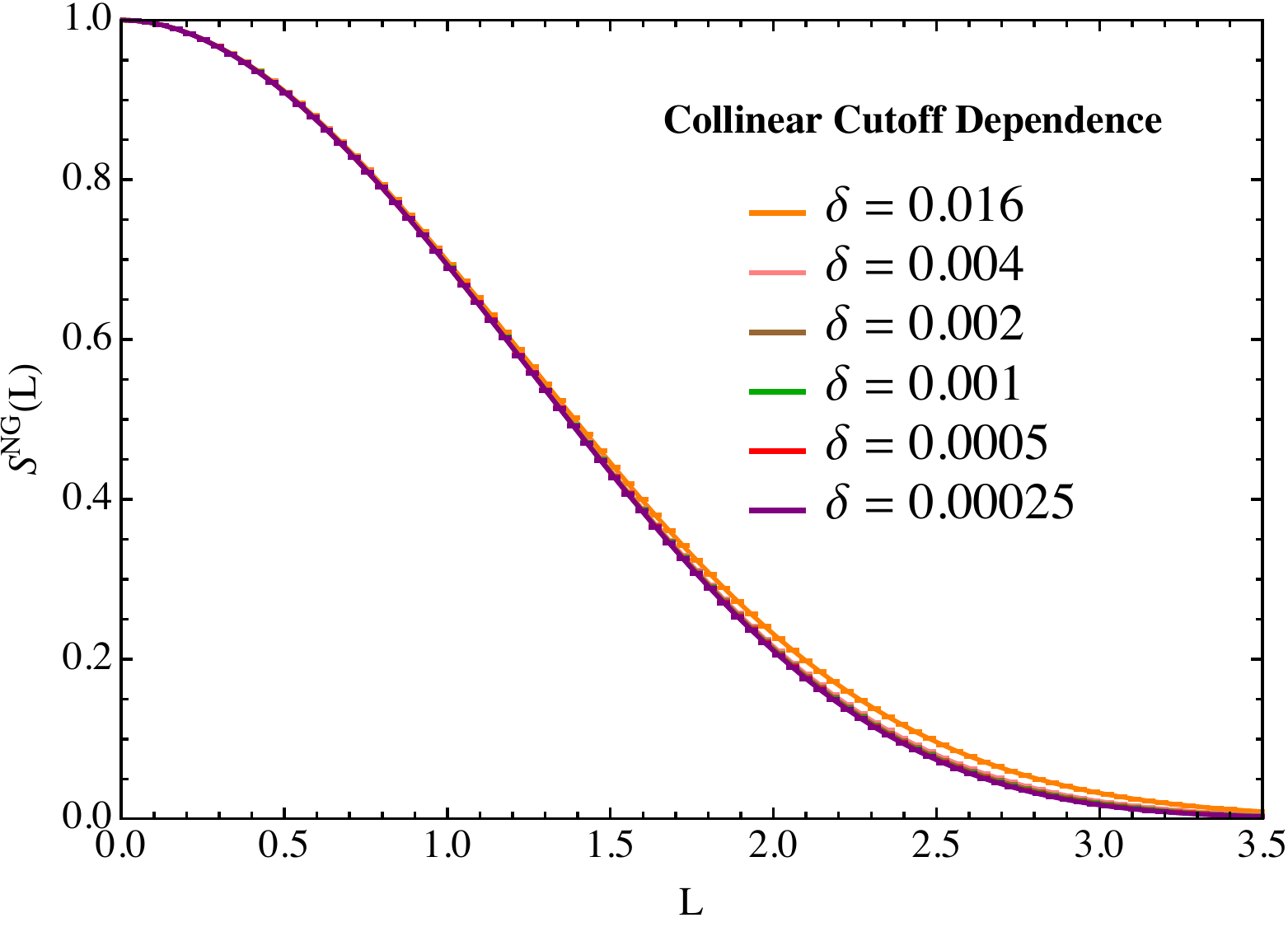}
}
\subfloat[]{\label{fig:b_t_b_Cutoff_b}
\includegraphics[width=7.3cm]{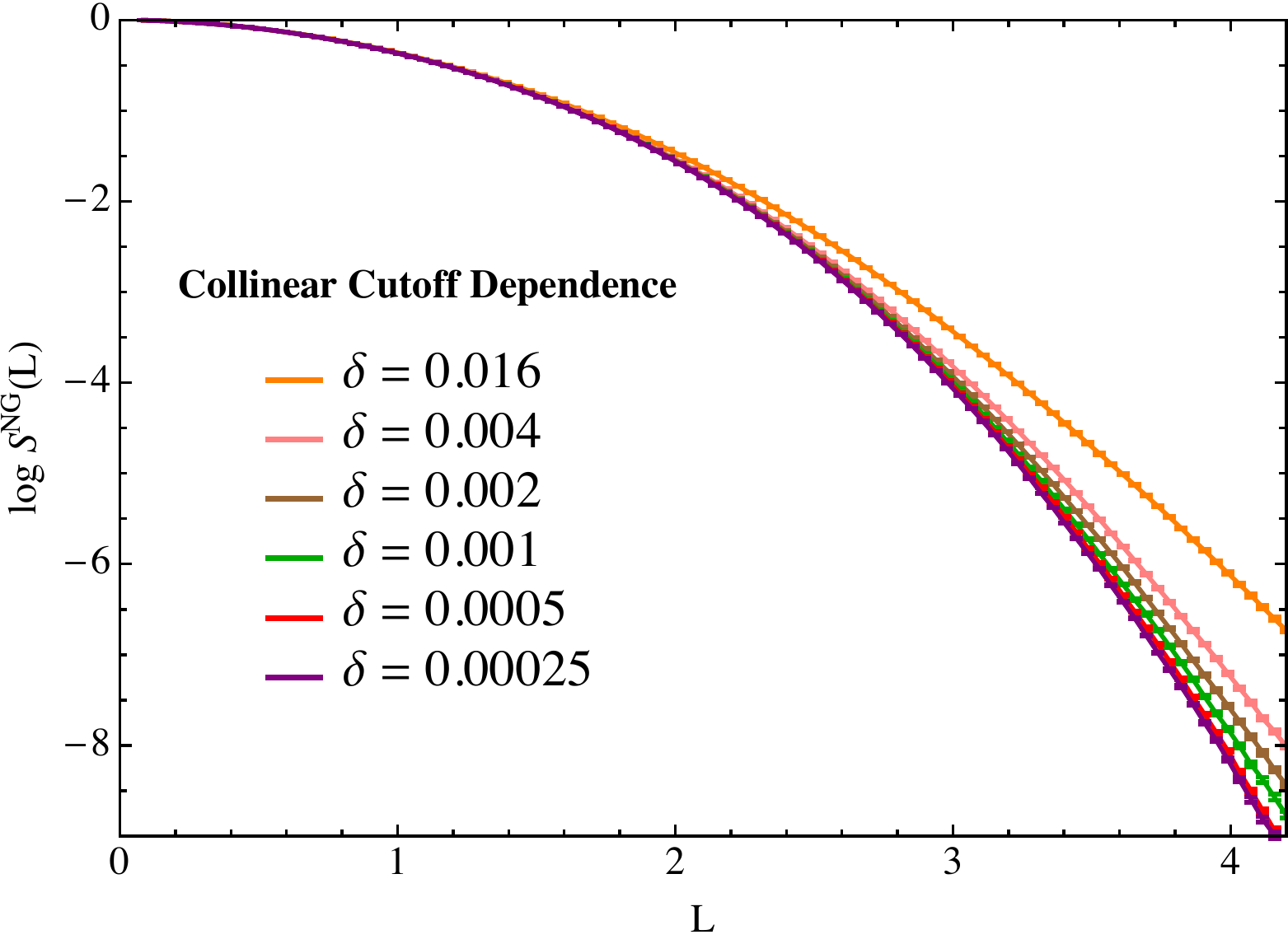}
}
\end{center}
\caption{The cutoff dependence of the Monte Carlo solution to the BMS equation. Displayed are both the distribution and the logarithm of the distribution, with statistical uncertainties. 
}
\label{fig:b_t_b_Cutoff}
\end{figure}

As an illustration of the effect of the collinear cutoff on the NGL distribution, we plot the large-$N_c$ Monte Carlo solution of \Ref{Dasgupta:2001sh} in a hemisphere dijet geometry in \Fig{fig:b_t_b_Cutoff}.\footnote{In general, we do not necessarily expect the spread of the Monte Carlo runs to be Gaussian distributed in each bin. Thus, to estimate statistical errors, we split the Monte Carlo into a hundred runs for each cut-off, where each run now has fewer events than the total collected. This gives us a distribution of runs for each bin, with which we can directly determine the fluctuation width of 65\% of the runs about the mean. This was found to be close to the root-mean-squared (RMS) analysis of distribution of runs. We also checked the same remained true when the Monte Carlo was divided into 50, 30, 25, and 15 runs, now with each run containing more events. Further, a random sampling of merely six runs gave a decent estimate of the total RMS. We then explicitly checked that the RMS estimate followed a $N^{-1/2}$ law as the number of events $N$ in each run was increased. For the final estimate of the statistical uncertainties, we took the RMS of the 100 run distribution, and rescaled it according to the $N^{-1/2}$ law to the total number of events collected. The total number events collected in each cut-off was greater than $4\times 10^6$.} The collinear cut-off in the Monte Carlo represents the smallest angle an emission can have in the event to another emission in the event in the lab frame. As is expected from the behavior of the lower bound of \Eq{eq:bounding_solutions_to_BMS}, the distribution decreases with decreasing collinear cutoff. The fact that the bounding solution is always a simple exponential, and goes to zero as $\delta\rightarrow 0$, indicates the behavior of the true distribution must decay more strongly than a simple exponential.

\section{Breakdown of the Fixed-Order Expansion for Non-Global Logarithms}\label{sec:fo_fail}

Having established the dressed gluon as a convergent expansion of the BMS solution,  we can use the structure of the dressed gluons to deduce properties about the full solution to the BMS equation. In particular, we show that the dressed gluons fulfill a necessary condition for the existence of a singular structure in the complex plane at $L=-1$ in the full solution of the BMS equation. If the BMS equation has such a singularity, and since the dressed gluon expansion converges everywhere to the solution, then each dressed gluon must exhibit singularities at the same point, in order for the singularities to not destroy the radius of convergence. In \Sec{sec:log_sing} we study the analytic behavior of the dressed gluons, showing that the single dressed gluon has a singularity at $L=-1$ in the complex plane, and therefore that its expansion in $\alpha_s$ has a finite radius of convergence. We then argue that this behavior persists for any number of dressed gluons. In \Sec{sec:LL_BMS_exp} we describe the physical origin of this singularity as being due to the buffer region, and explicitly show how this leads to coefficients of $\mathcal{O}(1)$ in the perturbative expansion of the dressed gluon. We then review how the dressed gluon naturally resums these contributions through its ``boundary soft" mode, leading to a convergent expansion. In \Sec{sec:coll_effect} we use the dressed gluon to study the behavior of the next-to-leading NGLs and show that collinear double logarithms worsen the convergence of the perturbative expansion if they are not resummed to all orders, as can be accomplished with the dressed gluons. This illustrates the importance of analysizing the factorization structure of the subjet production cross-sections.

\subsection{Logarithmic Singularities in the NGL Distribution}\label{sec:log_sing}

In this section, we will consider the analytic structure of a single dressed gluon off of the $n\nbar$ dipole with hemispherical jet regions, which we will already find to be quite interesting.  This configuration has been studied in detail and high-order perturbative results exists to which we can compare. We have from \Ref{Larkoski:2015zka}
\begin{align}\label{eq:1dressed}
\int_0^{L}dL'\,d_{n\nbar}(L')&=-\frac{\gamma_E}{2}L-\frac{1}{2}\,\log\, \Gamma(1+L)\,.
\end{align}
From this expression, one notes that there is a logarithmic singularity at $L=-1$. In particular, this implies that the fixed order expansion of the dressed gluon has a radius of convergence of $L=1$. Indeed, the Taylor expansion of the logarithmic factor entering the expression for the single dressed gluon is found to be
\begin{align}
-\frac{\gamma_E}{2}L-\frac{1}{2}\,\log\, \Gamma(1+L)=-\frac{1}{2}\sum_{i=2}^\infty\frac{\psi^{(i-1)}(1)}{i (i-1)!}L^i\,,
\end{align}
where $\psi^{(i)}(z)$ is the poly-gamma function
\begin{equation}
\psi^{(i)}(z) = \frac{d^{i+1}}{dz^{i+1}}\log \, \Gamma(z)\,.
\end{equation}
For large $i$,
\begin{equation}
\frac{\psi^{(i-1)}(1)}{i(i-1)!} \sim \mathcal{O}(1)\,,
\end{equation}
which explicitly exhibits this finite radius of convergence. This fact alone is quite remarkable, and shows that the dressed gluon expansion is capturing physics that is not described in fixed order perturbation theory to any order. In \Sec{sec:LL_BMS_exp} we will discuss the physics behind the finite radius of convergence for the expansion of the single dressed gluon in more detail, and how it is cured by the resummation included in the dressed gluon. Here we will show that this singularity arises mathematically from the behavior of $U_{n\bar nj}$ factor as $L$ becomes negative.  At negative $L$, the $U_{n\bar nj}$ factor no longer vanishes at the jet boundary in a power law fashion, suppressing emissions, but diverges instead. This leads not to a buffer region, but to a region with an unregulated number of emissions.\footnote{Amusingly, this flip of suppression and enhancement of emissions as the sign of $L$ is flipped is reminiscent of Dyson's argument for the divergence of the perturbative expansion in QED \cite{Dyson:1952tj}. However, in this case the divergence does not become sufficiently bad until $L=-1$, leading to a finite radius of convergence instead of zero radius of convergence.} We then generalize this argument to the $k$th dressed gluon.\footnote{Though we restrict ourselves here to back-to-back hemisphere jets, the logarithm of the angle of the soft-jet to the jet boundary is a generic feature of \emph{any} jet region. This follows from the fact that boundary soft modes which are introduced to resum this angle enjoy a collinear factorization, sensitive to the angular distance to the jet boundary. Indeed, one can check from calculations in \Ref{Ellis:2010rwa}, that these logarithms are present in a ``swiss cheese'' region jet region with arbitrary hard jets, and that they factorize collinearly from the soft color structure.}

The integral expression for the derivative of the first dressed gluon for an arbitrary initial dipole $ab$ is
\begin{align}\label{eq:dressed_gluon_integral_expression}
\partial_Lg^{(1)}_{ab}(L)&=\int_J\frac{d\Omega_j}{4\pi}W_{ab}(j)\left(U_{abj}(L)-1\right)\,.
\end{align}
Recalling the form of $U_{abj}(L)$ in \Eq{eq:NGL_anom_dim} for the hemisphere case, we can write as the emission approaches the jet boundary, $j\rightarrow \partial J$, for an arbitrary initial dipole $ab$
\begin{align}\label{eq:U_near_jet_boundary}
U_{abj}(L)&\sim \left(\frac{\pi}{2}-\theta_j\right)^{L} f_{abj}(L)\,.
\end{align}
Here $f_{abj}$ is a smooth and finite function of $j$ throughout the jet region $J$ for all $L$. As long as $L>-1$, the integral over the jet region can be performed; otherwise, we have an unbounded result. As $L\to -1$  in the complex plane, the boundary of the jet will dominate the integration in Eq. \eqref{eq:dressed_gluon_integral_expression}, and we can deduce the nature of the singularity by taking a small region around the jet boundary,\footnote{In particular, the region must be within the buffer region, excluding the initial dipole points.} and integrating. The rest of the jet region will give a subleading result. In the integrand,  we can set $\theta_j=\pi/2$, except for the singular behavior in Eq. \eqref{eq:U_near_jet_boundary}, and integrate over $\phi_j$ and $\theta_j$ to find
\begin{align}
\partial_Lg^{(1)}_{ab}=\frac{c}{L+1}+\mathcal{O}(1),\, L\rightarrow -1\,,
\end{align}
$c$ is some constant. We then integrate over $L$ to conclude that there will be generically a logarithmic singularity at $L=-1$ for the first dressed gluon off any dipole, namely
\begin{align}
g^{(1)}_{ab}\sim \log(L+1)+\mathcal{O}(1),\, L\rightarrow -1\,.
\end{align}

While we cannot compute the higher dressed gluon contributions analytically, the recursive nature of the dressed gluon expansion, \Eq{eq:DG_recursive}, naturally sets up an inductive argument.  It is straightforward to show, due to the non-linear term in the Picard iteration of \Eq{eq:DG_recursive}, that
\begin{align}
g_{ab}^{(k)}\sim \log^{2^k-1}(L+1)+\cdots,\, L\rightarrow -1\,,
\end{align}
where the dots represent subleading logarithms in this limit. 
Therefore, at each order, the $k$-th dressed gluon exhibits a logarithmic singularity to the $2^k-1$-th power as $L\to -1$ in the complex plane.  

\begin{figure}
\begin{center}
\subfloat[]{\label{fig:fixed_order_expansion}
\includegraphics[width=6.75cm]{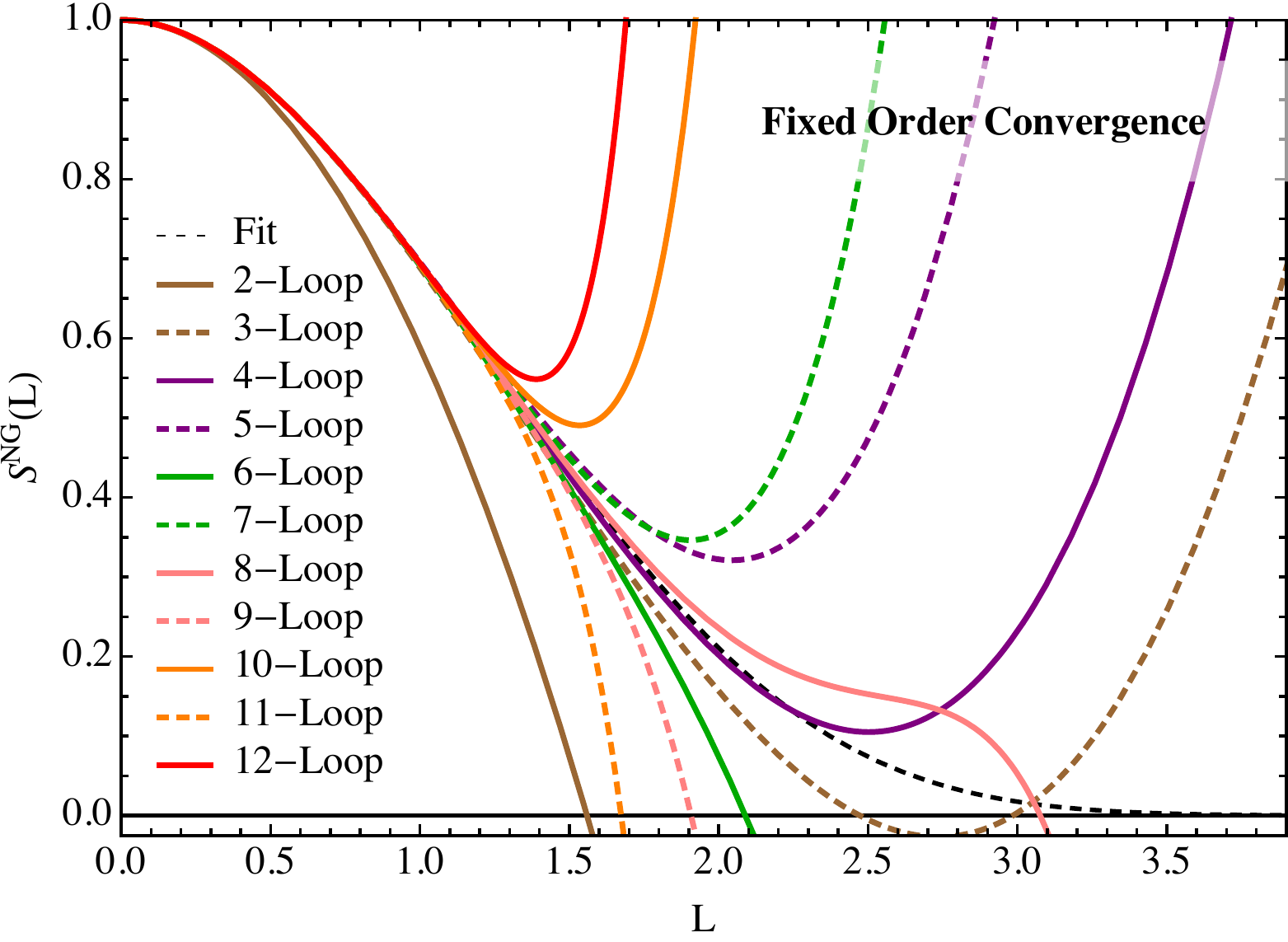}
}
\subfloat[]{\label{fig:fixed_order_expansion_ratio}
\includegraphics[width=7.15cm]{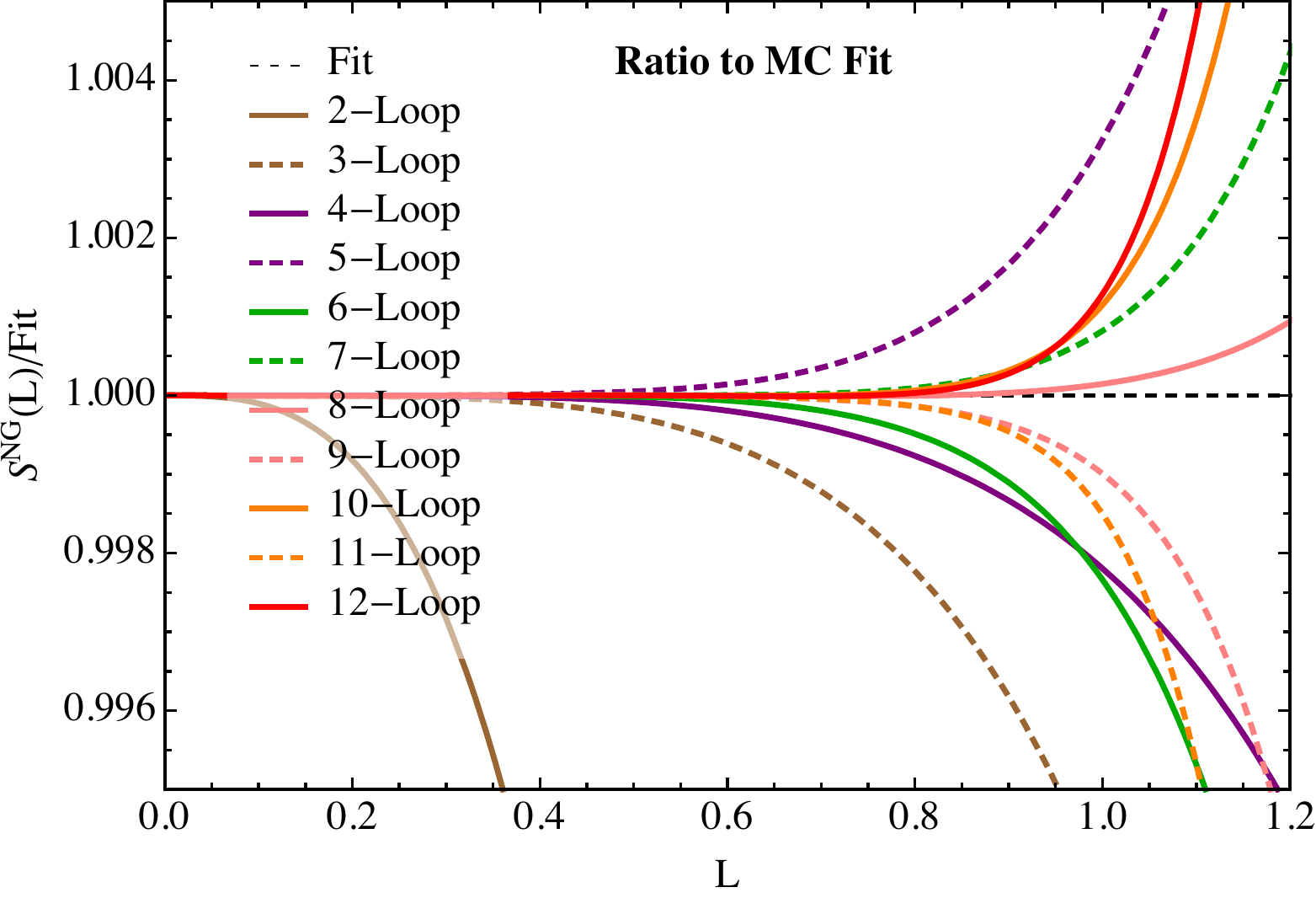}
}
\end{center}
\caption{The convergence of the fixed order expansion of the BMS equation for hemisphere mass to 12 loops. In (a) we show the distribution, and in (b), the ratio to the Monte Carlo fit. Even orders in the expansion are solid, odd orders are dashed. The behavior of the series is suggestive of a series with radius of convergence $L=1$.
}
\label{fig:expansions_fo}
\end{figure}

 It is important to emphasize that the presence of these logarithmic singularities does not effect the radius of convergence of BMS solution in terms of the dressed gluons themselves, since each dressed gluon itself has the logarithmic singularity fully constructed, as is necessarily the case if the full BMS solution has such a singularity. Indeed, this is what allows for the dressed gluons to provide an expansion with an infinite radius of convergence. Effectively, one is not expanding in $L$ near the logarithmic singularity, as is happening with the fixed order series, but in $\log(1+L)$. This is a highly non-trivial rearrangement of the standard fixed order expansion. Indeed, this is one of the advantages of the more general method of successive approximations, as compared with a more standard fixed order expansion.

\begin{figure}
\begin{center}
\includegraphics[width=8.5cm]{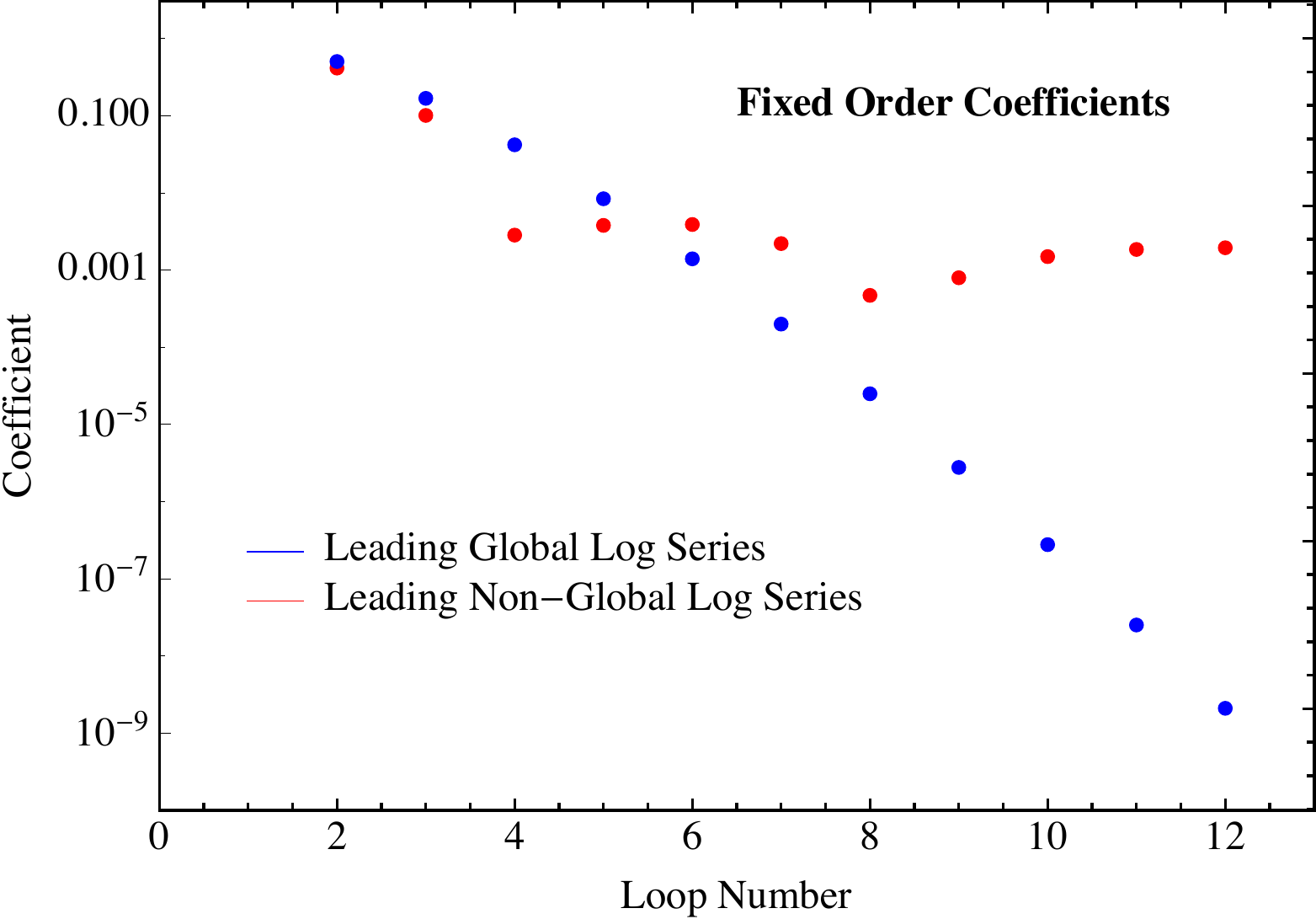}
\end{center}
\caption{The absolute value of the coefficients of the leading logarithmic, fixed-order expansion as a function of the order of the perturbative expansion for both global and non-global logarithms. The constant magnitude of the coefficient for the case of non-global logarithms supports a radius of convergence of the series of $L=1$. 
}
\label{fig:NGL_coeffs}
\end{figure}

Given these singularities, and the fact the dressed gluons converge to the full solution, we expect that the radius of convergence of the fixed order expansion of the BMS equation is $|L|=1$. As numerical evidence of this claim, in \Figs{fig:fixed_order_expansion}{fig:fixed_order_expansion_ratio} we plot the fixed order expansion of the leading order BMS equation to twelve loops for hemisphere jet mass \cite{SimonBMS}. In the ratio to the Monte-Carlo solution to the BMS equation, one can clearly see that regardless of the number of terms included in the fixed order expansion, the leading logarithmic series begin to diverge from the Monte Carlo solution at $L=1$. Thus while the series converges nicely for $L\leq 1$, above this value, the fixed order expansion does not describe the NGL distribution. Furthermore, in \Fig{fig:NGL_coeffs} we plot the absolute value of the coefficient of the leading logarithms up to 12 loops, and compare with that for the leading global logarithm, which has a infinite radius of convergence. For a general series with radius of convergence of $R$, the coefficients of the expansion, denoted here as $c_n$, obey
\begin{align}
\lim_{n\to \infty} \frac{c_{n+1}}{c_n}= \frac{1}{R}\,.
\end{align}
Unlike the coefficients of the global logarithmic series, up to 12 loops, the coefficients of the leading non-global logarithms are quite flat, supporting that the series indeed has a radius of convergence of $L=1$. Since this is a system where high perturbative orders can be computed, it would of course be interesting to test this to higher orders in the perturbative expansion.

\subsection{The Role of the Buffer Region and Boundary Soft Resummation}\label{sec:LL_BMS_exp}

It is enlightening to examine exactly how the buffer region is responsible for the finite radius of convergence of the fixed-order expansion.
Studying the buffer region will allow us to quantify how the coefficients of the fixed-order expansion grow as the number of loops increases, and highlights the contribution from logarithms sensitive to the angle to the jet boundary of the emission which are resummed in the dressed gluon approach. Since the angle of the dressed gluon is integrated over, these logarithms do not appear in the final result. Nevertheless, we show that they drive the breakdown of convergence of the fixed order expansion by contributing large coefficients to the perturbative expansion at each order.

As in the previous section, we again consider the behavior of the $U_{abj}$ factor as an emission approaches the boundary of the jet. We restrict to hemisphere jets, in this case $U_{abj}$ is exactly known, and is given in \App{sec:Uabj_hemisphere}, and further take $a=n, b=\nbar$. We have
\begin{align}
U_{n\bar nj}= \Big(1-\tan ^2\frac{\theta_j}{2}\Big)^{L}\,.
\end{align}
In a perturbative expansion in $\alpha_s$ (which we recall is absorbed into $L$), this term is expanded as
\begin{align}\label{eq:Uabj_expand}
\Big(1-\tan ^2\frac{\theta_j}{2}\Big)^{L}=\sum_{i=0}^\infty \log^i \Big(1-\tan ^2\frac{\theta_j}{2}\Big) \frac{L^i}{i!}
\end{align}
Importantly, this expansion is performed before integration over the phase space region of the jet. In this expansion, one sees logarithms of the angle of the dressed gluon (or subjet) to the jet boundary, that appear at each order in perturbation theory.

An important feature in the construction of the dressed gluon that was emphasized in \Ref{Larkoski:2015zka}, and reviewed in \Sec{sec:dressed_gluon_review}, is the ``boundary soft" mode which achieves a resummation of these logarithms. This was also required in the analytic calculation of the $D_2$ observable \cite{Larkoski:2015kga}. The size of the logarithms depends on the location in the phase space; however, these logarithms are large in the region of phase space that contributes to the NGLs, namely
\begin{align}
\log \Big(1-\tan ^2\frac{\theta_j}{2}\Big)\rightarrow \infty \text{ as } \theta_j\rightarrow \frac{\pi}{2}\,.
\end{align}

The manifestation of these logarithms in the final NGL series, is however, more subtle. Indeed, performing the marginalization over the dressed gluon phase space, one finds that these logarithms are integrable
\begin{align}\label{eq:logint}
\int^{0}dx\,\log^i x\sim i!\,,
\end{align}
leading not to a large logarithm, but to a large constant. However, it is precisely the factorial growth of these terms that will lead to a finite radius of convergence of the $\alpha_s$ expansion of the dressed gluons. Note that this is independent of the other endpoint of the integral (at sufficiently large $i$), and only depends on the fact that one must integrate to the boundary of the jet, which is the origin of the logarithmic divergence.\footnote{From our working definition of the buffer region, \Eq{eq:buffer_definition}, one can see the lower limit of the polar integration is set by the leg of the dipole closest to the boundary, since the resummation factor $U$ always goes to 1 at this leg. The closer this leg is to the boundary, the higher loop orders we expect one must compute before the coefficients of the NGL series flatten out. We have checked in other dipole configurations that the fixed order series does appear to lose convergence after $L=1$.}

To see this, we examine the fixed order expansion of $g_{ab}^{(1)}(L)$, the first dressed gluon
\begin{align}\label{eq:fixed_order_expansion_1DG_LL}
g^{(1)}_{ab}(L)=\sum_{i=0}^{\infty}d_i L^i\,.
\end{align}
Higher order emissions come with their own resummation factor $U_{abj}$, and it is straightforward to extend the argument to such terms. If we expand the $U_{abj}$ factor in the dressed gluon perturbatively before performing the integration, as in \Eq{eq:Uabj_expand},
then we have, using \Eq{eq:logint},
\begin{align}
d_i \sim \int_J\frac{d\Omega_j}{4\pi}W_{ab}(j)  \frac{ \log \Big(1-\tan ^2\frac{\theta_j}{2}\Big)^i}{i!}\sim 1\,,
\end{align}
since all other terms are order $1$ throughout the jet region in the anomalous dimension, and the eikonal factor is also generically order $1$ at the jet boundary. 
We therefore conclude that in \Eq{eq:fixed_order_expansion_1DG_LL}
\begin{align}
d_i\sim \mathcal{O}(1)\,,\forall\,i\,,
\end{align}
which implies a radius of convergence of $L=1$. It is important to emphasize that the behavior of the integral comes only from the endpoint of the integral, namely the boundary of the jet.

This clearly demonstrates the necessity of the resummation of the logarithms associated with the boundary of the jet for achieving a convergent perturbative expansion, as is achieved by the boundary soft modes in the dressed gluon expansion. It also emphasizes subtleties in marginalizing over factorization theorems, and the resummation of logarithms before and after marginalization.\footnote{In the approach of \Refs{Becher:2015hka,Becher:2016mmh} these logarithms are not resummed before marginalization.}

\subsection{Collinear Effects at Next-to-Leading Logarithm}\label{sec:coll_effect}

While we have so far focused in this paper on the leading logarithmic behavior of the NGL series, as described by the BMS equation and for which high loop perturbative data exists, it is also interesting to consider what happens at next-to-leading logarithmic (NLL) accuracy. Since the dressed gluon is described by an all-orders factorization theorem, it is systematically improvable, and allows such questions to be studied. The extension of the dressed gluon to NLL was discussed in \Ref{Neill:2015nya}.

At next-to-leading logarithmic order, collinear double logarithms become an important feature of the NGL series.\footnote{Such logarithms also have been argued to play an important role in the context of the BFKL \cite{Salam:1998tj,Ciafaloni:1999yw,Ciafaloni:2003rd,Vera:2005jt} and BK \cite{Iancu:2015vea,Iancu:2015joa} equations.} Again, their resummation can be achieved using the dressed gluon expansion. 
In this case, one must include not only the boundary soft modes, but another mode which is also sensitive to the boundary of the jet, referred to as the ``edge of jet" mode in \Ref{Neill:2015nya}.
Ultimately, the effect of these modes is to modify the $U_{abj}$ resummation factors in \Eq{eq:BMS_eqn_large_N}, introducing another evolution kernel
\begin{align}
U_{abj}\rightarrow U^{\text{ci}}_{abj}=U_{abj}U_{abj}^{\mathcal{E}}\,. 
\end{align} 
Here the ``ci" superscript stands for ``collinearly improved" and ${\mathcal{E}}$ denotes the evolution kernel for the edge of jet mode. 
For our purposes here, we can simply expand out this resummation factor to find the impact on the NLL series.  To the lowest orders for hemisphere jets, we schematically have (see \App{app:ci_of_U} for relevant evolution equations and scales)
\begin{align}\label{eq:expanding_running_coupling}
\log  \,U_{abj}^{\text{ci}}(m_H,m_L)&=\gamma_{abj}L -\frac{\alpha_s C_A}{4\pi}\beta_0\gamma_{abj}L\left(\log \left(\frac{m_H m_L\Delta\theta_j^2}{Q^2}\right)+3\gamma_{abj}\right)+\cdots \,, 
\end{align}
where we have suppressed the scale at which $\alpha_s$ is evaluated.  Here, $Q$ is the center of mass energy of the event, $\beta_0$ is the one-loop coefficient of the $\beta$-function, $\Delta\theta_j=\frac{\pi}{2}-\theta_j$ is the angle of the soft subjet to the jet boundary, and $\gamma_{abj}$ is the one-loop anomalous dimension of the boundary soft mode, proportional to the logarithm of the angle to the jet boundary.  For hemisphere jets with back-to-back dipoles,
\begin{equation}
\gamma_{n\bar n j} = \log\left(
1-\tan^2\frac{\theta_j}{2}
\right)\,.
\end{equation}
 Additionally, we generically expect that there is observable dependence in the anomalous dimension $\gamma_{abj}$ at higher orders, but here we ignore it for simplicity.

Note that in this expression there is a genuine double logarithm of $\Delta \theta_j=\pi/2-\theta_j$.  To determine the evolution equation that includes splittings along the boundary accurate to NLL, we must expand the kernel $U_{abj}^{\text{ci}}$.  From the expansion in \Eq{eq:expanding_running_coupling}, we have
\begin{align}
U_{abj}^{\text{ci}}(m_H,m_L)
&\simeq \exp\left[
\gamma_{abj}L -\frac{\alpha_s C_A}{4\pi}\beta_0\gamma_{abj}L\left(\log \left(\frac{m_H m_L\Delta\theta_j^2}{Q^2}\right)+3\gamma_{abj}\right)
\right] \\
&
=\sum_{n=0}^\infty \frac{1}{n!}\left[
\gamma_{abj}L -\frac{\alpha_s C_A}{4\pi}\beta_0\gamma_{abj}L\left(\log \left(\frac{m_H m_L\Delta\theta_j^2}{Q^2}\right)+3\gamma_{abj}\right)
\right]^n \nn \\
&
\simeq\sum_{n=0}^\infty \frac{(\gamma_{abj}L)^n}{n!}\left[
1 -n\frac{\alpha_s C_A}{4\pi}\beta_0\left(\log \left(\frac{m_H m_L\Delta\theta_j^2}{Q^2}\right)+3\gamma_{abj}\right)
\right] \nn \,.
\end{align} 
Then, with the edge of jet modes included, in the fixed order expansion of the NGLs, we will encounter integrals of the form
\begin{align}\label{eq:coll_series}
\frac{1}{n!}\cdot n\cdot\int_J\frac{d\Omega_j}{4\pi}W_{ab}(j)\log ^{n+1}\Delta\theta_j\sim n^2\,,
\end{align}
and which therefore have a worse behavior than at LL.
By the ratio test for series convergence, one still expects an $\mathcal{O}(1)$ region of convergence; however, the coefficients of the NLL series at each loop order are now growing quadratically. Indeed, it is straightforward to see that these collinear logarithms at higher subleading logarithmic order (NNLL, N$^3$LL, etc.) will eventually drive a factorial growth in the coefficients of the fixed order perturbation series if they are not resummed. 

It is interesting to understand the role of the running coupling in generating these contributions. To get the factorial growth, it is enough to truncate the running of the coupling to its first perturbative order as in \Eq{eq:expanding_running_coupling}. That is, the factorial growth is not due to the all-orders running of the coupling probing a renormalon, as is the usual bubble chain analysis; see \Ref{Beneke:1998ui}, but simply the chain of collinear splittings along the boundary. The NLL series is significantly more complicated to calculate, as it cannot be computed in the strong energy ordered limit. It would however, be extremely interesting to verify the prediction of \Eq{eq:coll_series} explicitly.

Due to the behavior of the NLL series, it is important to emphasize how this shows that the standard logarithmic organization of the perturbative series that is traditionally used for global observables is not appropriate for the case of NGLs. The standard counting of logarithms used for global observables assumes that the coefficients of the logarithms (neglecting the factors of $\alpha_s$) are $\mathcal{O}(1)$ numbers. In this case, there is a suppression between LL and NLL due to the additional power of $\alpha_s$. However, in the NGL case, this suppression by a factor of $\alpha_s$ at NLL is accompanied by a quadratic growth of the coefficients, which will overwhelm the suppression by $\alpha_s$ at a certain loop order. This implies that the standard logarithmic counting used for global observables is not appropriate, and that these effects due to the buffer region and collinear divergences associated with the edge of the jet must be resummed to all orders to have a convergent perturbative expansion. This is naturally achieved by the dressed gluon expansion.

As a further comment, one might expect a cancellation of the collinear resummation, since na\"ively, the resummation of NGLs should be driven by soft physics. The collinear splittings in the heavy hemisphere are at the energy scale $m_H$ and are required to always remain within that hemisphere, and thus are sensitive to the scale $\Delta\theta_j$. Collinear splittings in the other hemisphere are at amuch lower scale, $m_L$, and can only approach to within $\Delta\theta_j$ of the soft subjet axis. Thus the collinear splitting angle is independent of the hemisphere, and because the corresponding anomalous dimensions have opposite sign, one might expect such a cancellation. Indeed, in a conformal theory, collinear splittings are not sensitive to the absolute energy of the parent parton, and so one would expect the collinear scale (the angle to the edge of the jet) to cancel between the in-jet and out-of-jet splittings, and one can adopt a scheme between the LO and NLO kernels such that such logarithms are canceled automatically in the evolution. Thus the coefficients of the NLL series would remain $\mathcal{O}(1)$ numbers in a conformal theory.

\subsection{Improving the Fixed Order Perturbative Convergence}\label{sec:conformal_improvements}

A number of techniques exist for improving the behavior of poorly convergent expansions, such as Borel resummation, Pad\'e approximants, or order-dependent mappings \cite{Seznec:1979ev}. Given an understanding of the analytic structure of the solution, one approach is to use conformal transformations to extend the domain of analyticity. This has been explored in the context of QCD in 
\Refs{Altarelli:1994vz,Caprini:2000js,Caprini:2010ir}; see also  \Ref{ZinnJustin:2002ru} for a pedagogical review.\footnote{We are grateful to Martin Beneke for suggesting this approach to us, and for directing us to the relevant literature.} Given our understanding of the analytic structure obtained from the study of the dressed gluon expansion, namely that we suspect that there is a branch cut singularity at $L=-1$, we can apply a conformal transformation \cite{ZinnJustin:2002ru}
\begin{align}\label{eq:conf_map}
L\rightarrow u(L)\,,
\end{align}
and derive from the original power series for the BMS equation in $L$ an improved power series in the variable $u$.\footnote{Note that this approach has typically been applied to improve the radius of convergence in the Borel plane for factorially divergent series. Since we have argued that the expansion of the BMS equation has a finite radius of convergence, we can directly apply the mapping to the series, and do not first apply a Borel transform.} The function $u$ must be a function that has a common domain of analyticity as the original power series in $L$. Put simply, $u$ must also have a power series at $L=0$ in terms of $L$, and thus within this domain defines a conformal mapping. We now write an expansion:
\begin{align}
g_{ab}(u)&=\sum_{i=0}^{\infty}c_{ab}^{(i)}u^i\,.
\end{align}
The coefficients $c_{ab}^{(i)}$ are fixed by requiring that the power series in $L$ is reproduced after substituting the explicit $L$ dependence in. Importantly, the new power series in $u$ can have a much larger domain of analyticity as a function of $u$ than the original power series in $L$. In general, we can go to the same order in the conformally mapped distribution as we can in the original series. We consider two such mappings:
\begin{align}
u(L)=\begin{cases}
               \frac{\sqrt{1+L}-1}{\sqrt{1+L}+1}\,,\\
               \log(1+L)\,.
 \end{cases}
\end{align}
The first is commonly used in renormalon analysis, and conformally maps the $L$ plane to a disc, whereas the second is motivated by the fact that the dressed gluon expansion explicitly indicates a series of logarithmic singularities at $L=-1$. Both have the feature that the singularity at $L=-1$ is pushed to the boundary of the domain of the mapping, far from $u=0$. We will refer to these as the disc mapping and the log mapping, respectively.\footnote{These mappings give rise to series which are obviously not series in $\alpha_s$, but are series in functions of $\alpha_s$. It is perhaps interesting that one of the early examples in jet physics where an expansion in a non-trivial function of $\alpha_s$ appeared was in the anomalous dimension for multiplicity, which is proportional to $\sqrt{\alpha_s}$ at lowest order \cite{Mueller:1981ex,Bassetto:1982ma,Mueller:1983cq}. As is well known (see e.g.~\Ref{Ellis:1991qj}) the anomalous dimension admits an expansion in $\alpha_s$, but with a radius of convergence of $|\alpha_s| < \frac{\pi}{8C_A}|j-1|^2$, where $j$ is the order of the Mellin moment. To extend the radius of convergence, one can perform a remapping (resummation) similar to those presented here, which leads to the $\sqrt{\alpha_s}$ behavior.}

\begin{figure}
\begin{center}
\subfloat[]{\label{fig:conformal_improvement_dg_a}
\includegraphics[width=7.5cm]{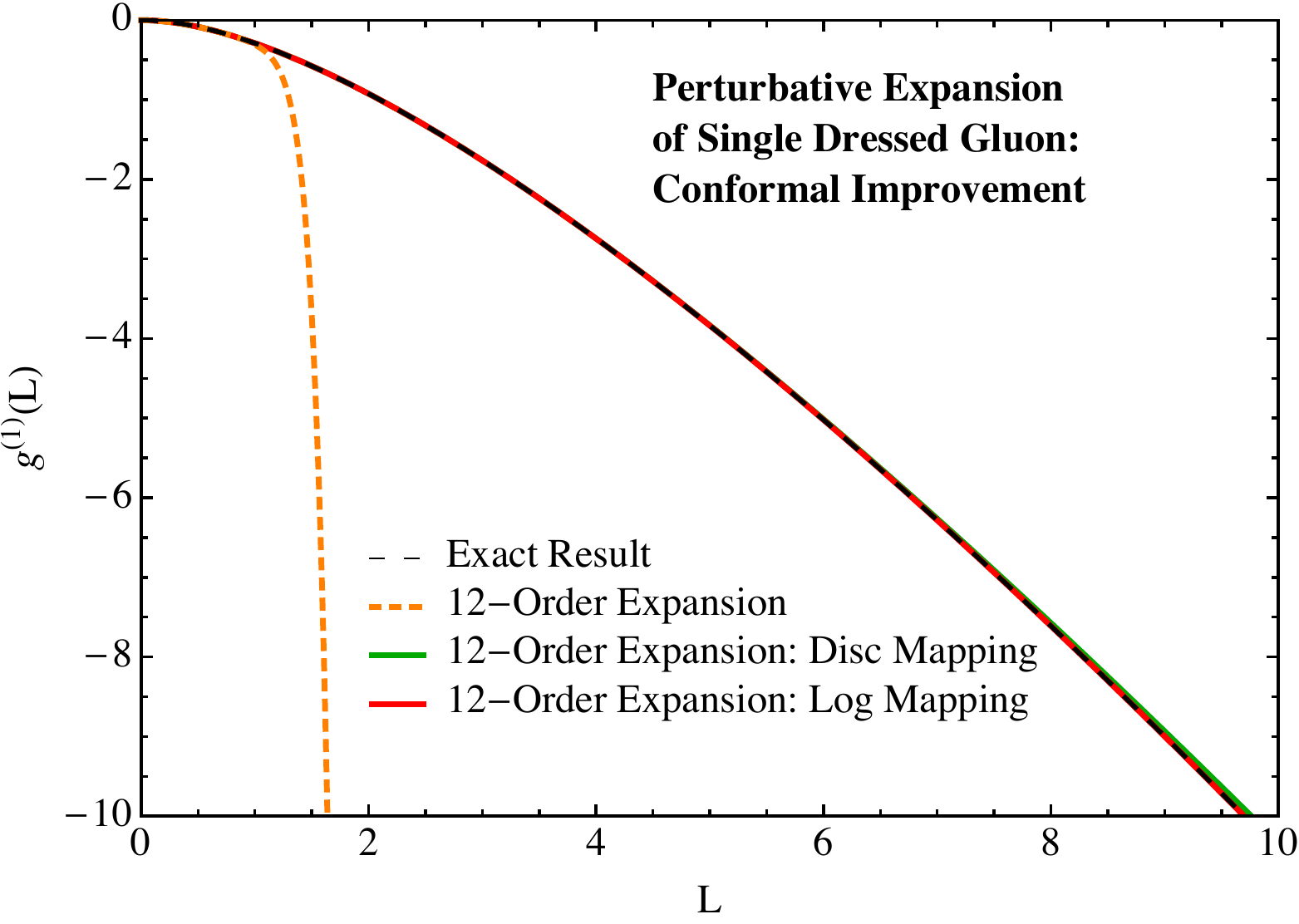}
}
\subfloat[]{\label{fig:conformal_improvement_dg_b}
\includegraphics[width=7.5cm]{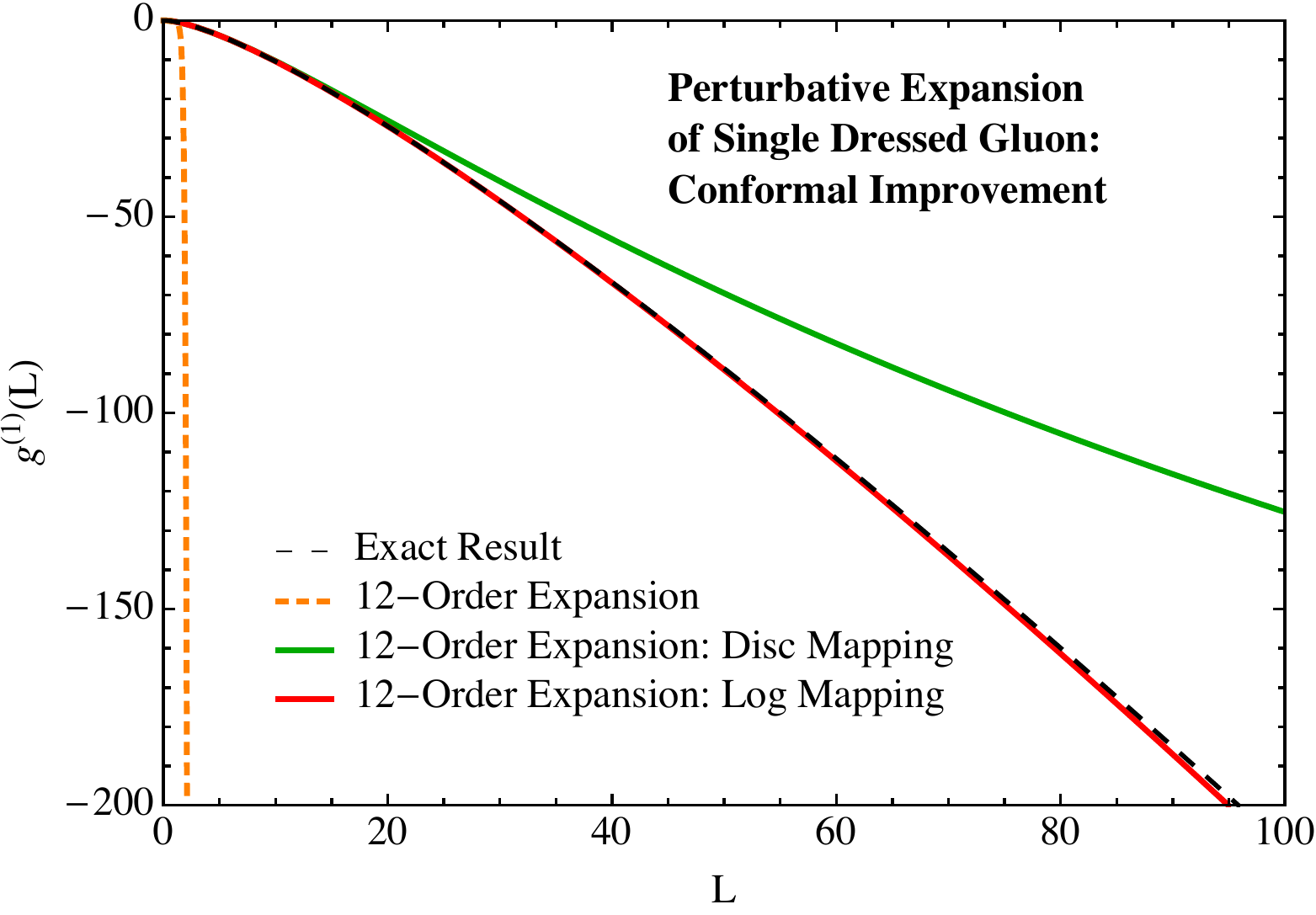}
}
\end{center}
\caption{The conformal improvement of the fixed order expansion of the single-dressed gluon using both the disc and log mappings described in the text. A zoomed in version is shown in (a), and a zoomed out version in (b). While the fixed order expansion has a radius of convergence of $L=1$, the conformally improved expansions have an infinite radius of convergence. The expansion using the log mapping is particularly fast.
}
\label{fig:conformal_improvement_dg}
\end{figure}

\begin{figure}
\begin{center}
\subfloat[]{\label{fig:NGL_Log}
\includegraphics[width=7.5cm]{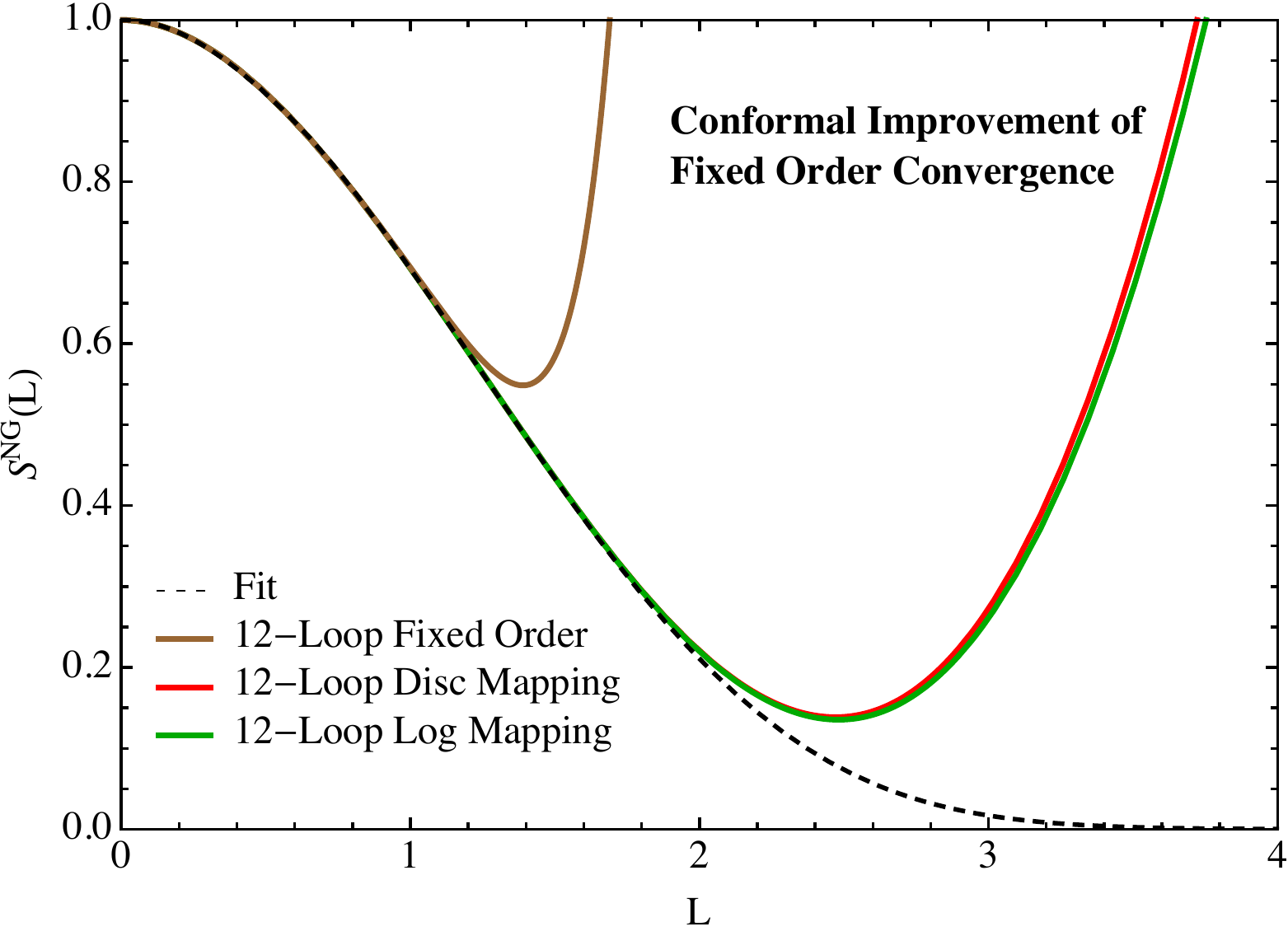}
}
\subfloat[]{\label{fig:NGL_Root}
\includegraphics[width=7.5cm]{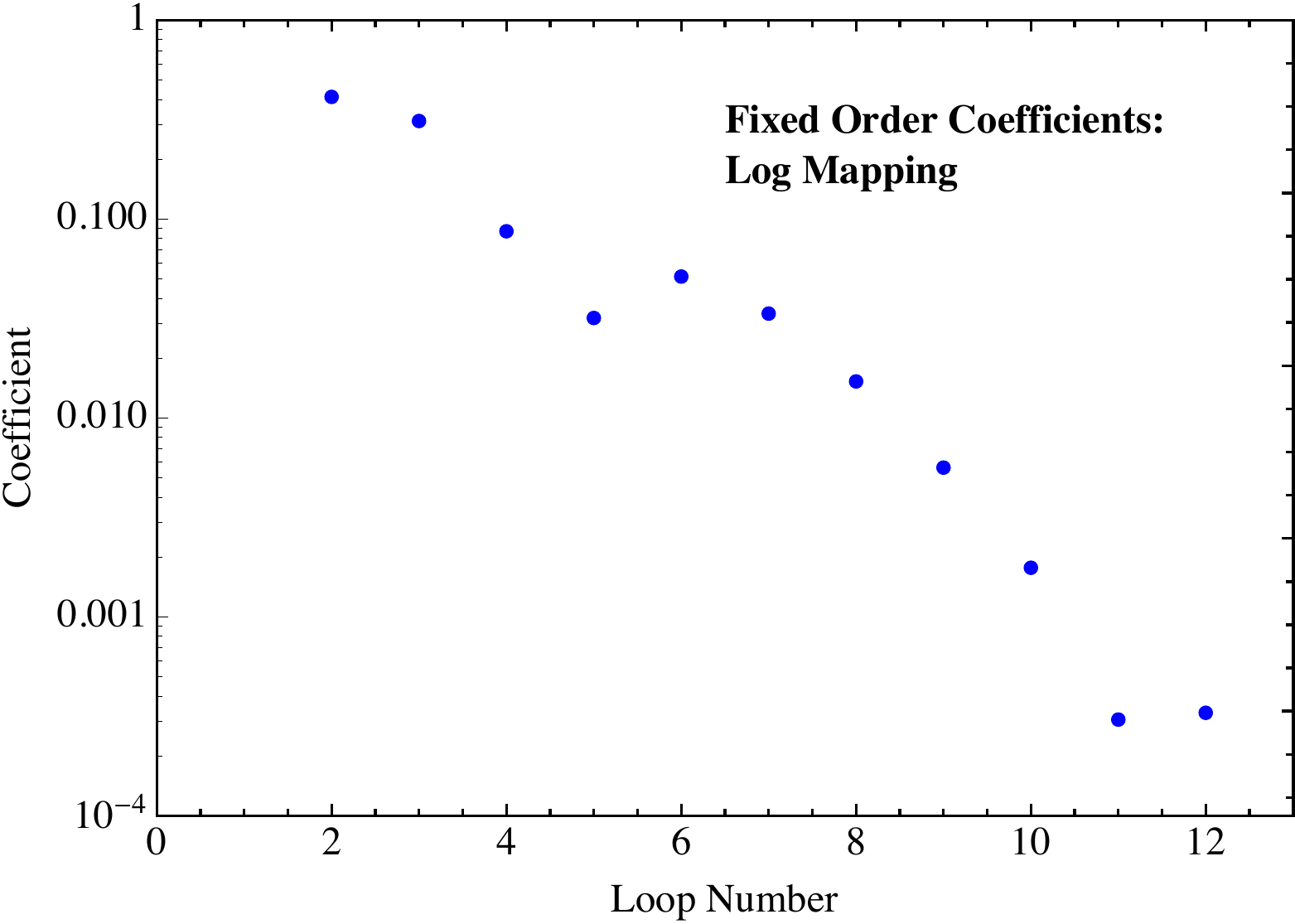}
}
\end{center}
\caption{A comparison of the $12$-loop fixed order expansion for the NGL distribution with the conformally improved fixed order expansion using both the disc mapping and the log mapping in (a). A significant improvement in the convergence is seen for both conformally improved series, well beyond $L=1$. In (b) we show the absolute values of the coefficients in the expansion for the log mapping.
}
\label{fig:conformal_improvement}
\end{figure}

We begin by testing these approaches on the single dressed gluon for back-to-back dipoles in the hemisphere mass case, for which the analytic expression is known (\Eq{eq:1dressed}). In \Fig{fig:conformal_improvement_dg} we show a comparison of the fixed order expansion, compared with the two different conformal improvements. Here we have chosen to work to $12$th order in the expansion, simply for illustrative purposes, as this is the order that the NGL series is known to. While the fixed order expansion has a radius of convergence of $L=1$, as illustrated by the divergence of the $12$th order expansion at this point, both conformal improvements have an infinite radius of convergence.  The two conformally improved series do illustrate different rates of convergence. For the log mapping, the convergence is remarkably fast, as is perhaps expected, as it was inspired by the form of the dressed gluon.

Given the perturbative expansion of the BMS equation, which is known to $12$-loops, we can test this resummation approach. In \Fig{fig:conformal_improvement} we show a comparison of the perturbative expansion at $12$-loop order, with the two conformally improved series. While the perturbative expansion diverges at $L\simeq 1$, the conformally improved solution, which was obtained only from the coefficients of the fixed order expansion, and the assumption of a branch cut at $L=-1$, exhibits considerably improved convergence. With the $12$-loop results, excellent agreement is seen out to $L\simeq 2$. This is well beyond the apparent radius of convergence of the fixed order expansion. Unlike for the case of the single dressed gluon, comparable convergence is seen for the two different mappings. This is perhaps not surprising. For the case of the single dressed gluon, we knew not only the presence of a branch cut, but that it was logarithmic. Much better convergence was then seen with the logarithmic mapping. However, for the full NGL series, from the arguments in \Sec{sec:log_sing}, we know that the nature of the singularity is more complicated. Both expansions we considered therefore only incorporate the location of the singularity.  This illustrates that having a better understanding of the analytic structure can lead to considerably more rapid convergence.

We emphasize that these conformal mappings are a form of resummation, similar in spirit to the explicit resummation associated with the dressed gluon, but here captured in a purely algebraic form. Just as in that case, it is this resummation which allows for a radius of convergence beyond that of the fixed order expansion. We expect that, like the dressed gluon expansion, the radius of convergence of the conformally improved NGL series is infinite, but we do not have a proof of that fact.

\begin{figure}
\begin{center}
\subfloat[]{\label{fig:conformal_improvement_II_a}
\includegraphics[width=7.5cm]{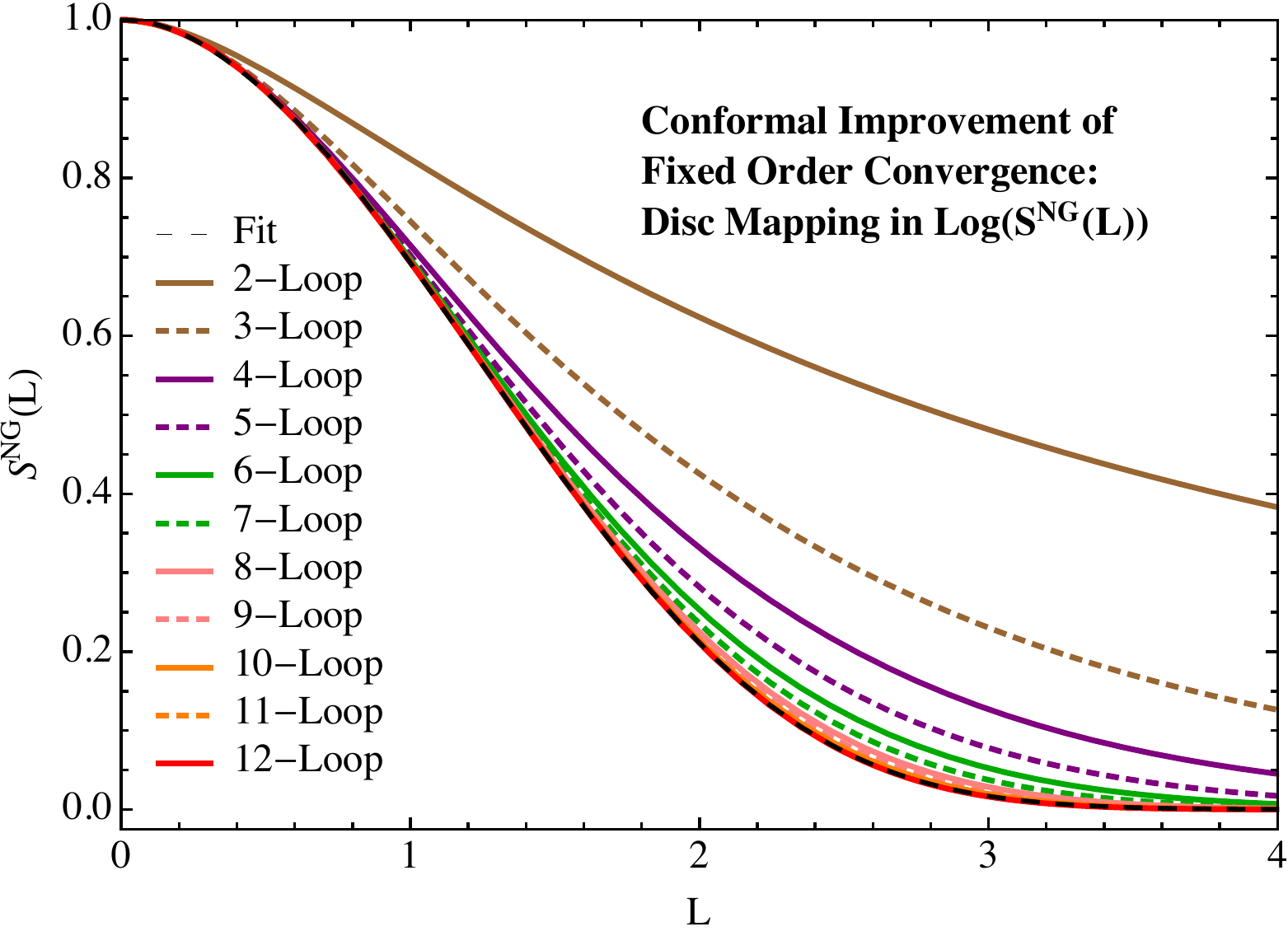}
}
\subfloat[]{\label{fig:conformal_improvement_II_b}
\includegraphics[width=7.75cm]{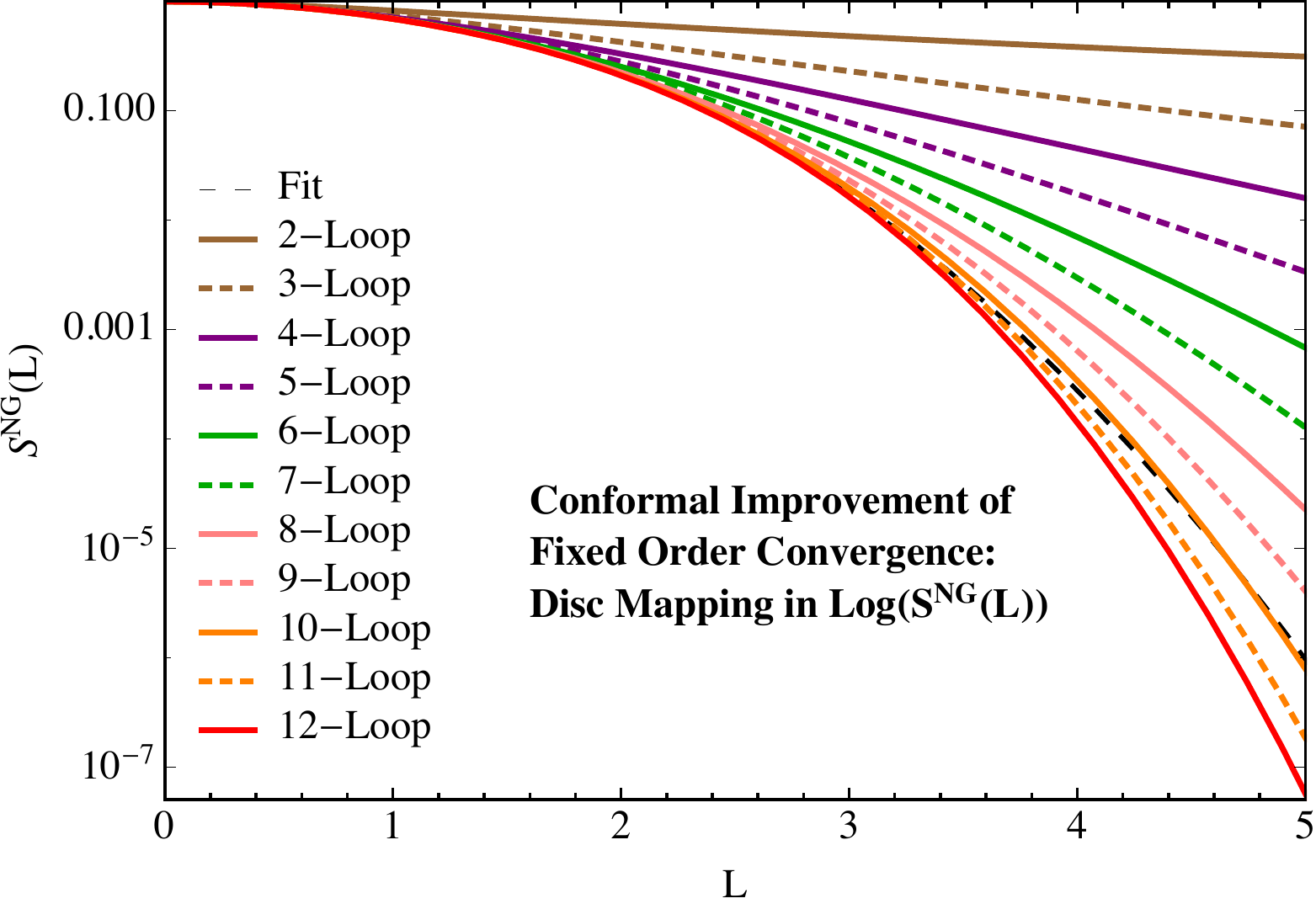}
}
\end{center}
\caption{The convergence of the conformally improved NGL series using a disc mapping in the logarithm of the distribution. Linear plot is shown in (a), and logarithmic plot in (b). Odd orders in the expansion are dashed, even orders solid. Excellent convergence is observed. The fit to the Monte Carlo should not be trusted beyond $L\sim 3$ due to the effect of a finite collinear regulator.
}
\label{fig:conformal_improvement_II}
\end{figure}

As an example of other possible mappings that can be used to improve convergence, we can also apply the conformal mapping approach to the logarithm of the NGL distribution. Based on our above arguments, the logarithm of the NGL distribution also has a finite radius of convergence, and therefore performing a strict fixed order expansion of the logarithm of the NGL distribution does not improve convergence. However, performing the conformal improvement in the logarithm of the distribution allows us to capture both the analytic structure on the negative real axis, as well as the behavior at large $L$. In \Fig{fig:conformal_improvement_II} we show the convergence of the NGL series using the conformal improvement of the logarithm of the distribution based on the disc mapping. Remarkably good convergence is seen for all $L$, and this convergence is extremely uniform. One should compare this with the pure fixed order expansion in \Fig{fig:expansions_fo}, which used the same color convention for the loop orders. The control over the series obtained using the conformal mapping is evident. 

In  \Fig{fig:conformal_improvement_II_b} the fit to the Monte Carlo should not be trusted beyond $L\sim 3$.\footnote{This particular value is identified by varying the collinear cutoff about the minimal value at which we ran the Monte Carlo.} Indeed, in this region the conformally improved result is slightly below the Monte Carlo fit, as would be expected from a finite collinear regulator (see \Fig{fig:b_t_b_Cutoff}), as was discussed in \Sec{sec:coll_reg}. From the behavior of the series, we can expect that the conformally improved result is more accurate than the Monte Carlo in this region. Under the assumption that this series converges, this allows us to get precise analytic predictions in the large $L$ region using fixed order perturbation theory, and not relying on Monte Carlo simulations with a finite collinear cutoff. This is not possible with fixed order perturbation theory alone, and requires the use of a resummation, as provided by the conformal mapping.

The ability to improve the perturbative convergence using a mapping based on a branch cut at $L=-1$ in the complex plane inferred from the structure of the dressed gluons provides further support that this feature is indeed present in the full BMS solution, as has been argued above, and we expect that the conformally improved expansion converges. It also emphasizes how insights into the analytic structure gained by studying the dressed gluon can be used to improve the perturbative understanding of NGLs. It would be interesting to study other approaches to reconstructing the full series from the perturbative expansion, as well as to study these expansions at higher orders.

\section{Conclusions}\label{sec:conc}

Non-global logarithms describe the entanglement due to soft emissions between distinct regions of phase space where different measurements are made. Unlike global logarithms, which exponentiate in a simple manner, non-global logarithms exhibit a much richer structure, being described by a non-linear integro-differential equation. This has made understanding their behavior, as well as performing their resummation, difficult. Recently, we proposed a reorganiation of the degrees of freedom that contribute to the NGLs, called the dressed gluon expansion. 

In this paper, we have elaborated on many aspects of the dressed gluon expansion, and used it to provide insights into the behavior of the NGL series. We gave a rigorous definition of the dressed gluon expansion at LL order by relating it to the method of successive approximations, and proved that it has an infinite radius of convergence as a solution to the BMS equation. This implies that it can be reliably truncated, and that its properties can be used to study the full solution of the BMS equation. The method of successive approximations is more general than a strict perturbative expansion, particularly in the case of non-linear equations, such as the BMS equation.

An interesting feature of the dressed gluon expansion is the analytic structure of the dressed gluons and the relation to the buffer region, a region of phase space near the boundary of the jet where emissions are suppressed. The single dressed gluon exhibits a singularity at $L=-1$ in the complex plane, and therefore, its fixed order expansion in $\alpha_s$ has a finite radius of convergence, namely $L=1$. In particular, this implies that the dressed gluon captures physics which cannot be calculated at any order in perturbation theory. We showed that this breakdown of the perturbative expansion is due to the dynamics of the buffer region, which contributes $\mathcal{O}(1)$ coefficients to the perturbative expansion. In the dressed gluon expansion, such contributions are resummed by the so called boundary soft mode, leading to a convergent series, which can be reliably truncated at each order. We have argued that such divergences are present in the full solution of the BMS equation, and we have studied the behavior of the known perturbative expansion of the BMS equation to 12-loop order, which exhibits the expected behavior for an expansion with radius of convergence $L=1$. We also discussed how at next-to-leading logarithm another class of contributions, arising from collinear splittings along the boundary, further spoil the perturbative convergence. These contributions can again be resummed using the dressed gluon expansion. 

Finally, we showed how an understanding of the analytic structure of the BMS solution, obtained using the dressed gluon expansion, can be used to improve the perturbative convergence of the fixed order expansion of the BMS equation using conformal mappings. This allows the use of fixed order perturbation theory to predict the distribution at large values of $L$, beyonds its na\"ive radius of convergence.

From a formal perspective, the structure of NGL series reveals a limit to fixed order perturbation theory in describing \emph{perturbative} QCD. The BMS equation and its resummation of NGLs was originally derived based on the analysis of the most singular region of the Feynman diagram expansion for the cross-section (see \Refs{Banfi:2002hw, Bassetto:1984ik}) using the recursive insertion of soft eikonal currents. It is important to emphasize the NGL distribution is collinear and infrared safe, so that non-perturbative effects can be considered as power corrections for much of the distribution. However, the BMS equation captures emergent dynamical behavior about jets, the buffer region, that in turn places a limitation on simply summing the Feynman diagram expansion to describe the distribution. This provides a precise definition of what is meant by emergent: in contrast to global logarithms, the resummation of NGLs through evolution equations, Monte Carlo simulation, the dressed gluon expansion, or conformal improvements of the series is not just helpful in stabilizing the perturbative result, but necessary to be able to make predictions for all values of the observable.\footnote{Similar claims can be made about the small-$x$ distribution of the parton distribution function; for instance the resummed DGLAP anomalous dimension in \Ref{Jaroszewicz:1982gr}. However, one must always deal with the infrared divergences of the initial state, and scheme dependence of the anomalous dimension, which can obscure making statements about the expansion of the cross-section. In this sense the NGL distribution is a much cleaner example.}

We believe that NGLs represent a fascinating playground for studying the perturbative structure of QCD. On the one hand, the all-orders result is described a known non-linear integro-differential equation, the BMS equation. On the other hand, the leading NGL series, being simpler than a generic cross-section, can be computed to high orders in perturbation theory, yet the coefficients exhibit similar structures to those found in $\mathcal{N}=4$ scattering amplitudes. Due to its convergence, the dressed gluon gives an analytic handle on the BMS equation, and the physics of non-global logarithms, which is not provided by the fixed order expansion. We hope that it will continue to provide a useful tool both for incorporating leading NGLs into factorization theorems for observables of phenomenological interest, as well as for studying the analytic structure of the solution of the BMS equation.

\acknowledgments

We thank Simon Caron-Huot for interesting discussions on the convergence of the fixed order series, and graciously providing us with the leading logarithmic NGL series through 12 loops. We thank Martin Beneke for suggesting the use of conformal mappings for improving the perturbative expansion of the NGL series, and for pointing us to relevant references on the subject. We would also like to thank Thomas Becher, Mrinal Dasgupta, Chris Lee, Matt Schwartz, Ding-Yu Shao, Iain Stewart, and Hua-Xing Zhu for helpful discussions as well as Anja Weyant for discussion of Monte Carlo uncertainties. This work is supported by the U.S. Department of Energy (DOE) under cooperative research agreements DE-FG02-05ER-41360, and DE-SC0011090, and also under contract DE-AC52-06NA25396 and through the LANL/LDRD Program.
A.L.~is supported by the U.S. National Science Foundation, under grant PHY--1419008, the LHC Theory Initiative. We also thank the Erwin Schrodinger Institute's program 
``Challenges and Concepts for Field Theory and Applications in the Era of the LHC Run-2", where portions of this work were completed.
\appendix

\section{$U_{abj}$ for Hemispherical Jets}\label{sec:Uabj_hemisphere}

For convenience, in this appendix we summarize the explicit form of the resummation factor $U_{abj}$ for the case of hemispherical jets, which was given in \Ref{Schwartz:2014wha}, and whose notation we follow. Due to the symmetries of the BMS equation, we need only $U_{anj}$ and $U_{abj}$, where $a,b,j$ go left. We will refer to these as the in-out and in-in case, respectively. We will write the expressions in terms of round and square bracketed inner products
\begin{align}
(ab)&=1-\cos \theta_{ab}=1-\cos \theta_a\cos \theta_b-\cos (\phi_a-\phi_b)\sin \theta_a\sin \theta_a\,,\\
[ab]&=(\bar {a} b)=1+\cos \theta_a\cos \theta_b-\cos (\phi_a-\phi_b)\sin \theta_a\sin \theta_a\,,
\end{align}
where here we have adopted the convention that $\theta_{ab}$ is the angle between the spatial components of the null vectors $a$ and $b$, and we also take $\bar{a}$ to be the reflection of the spatial components of $a$ into the other hemisphere.

For the in-in case, we have
\begin{align}
U_{abj}(L)=2^{L/2}\cos^L \theta_j \left(  \frac{[ab]}{[aj][jb]}  \right)^{L/2}\,,
\end{align}
where $\theta_j$ is the angle of the soft jet $j$ to the jet axis. We see explicitly that the $\cos^L \theta_j$ factor reproduces the behavior of the buffer region discussed in the text. Similarly, for the in-out case, we have
\begin{align}
U_{anj}(L)=2^{L/2}\cos^L \theta_j \left(  \frac{(an)}{[aj](jn)}  \right)^{L/2}\,.
\end{align}
In the text, we also make use of $\gamma_{abj}$, defined by
\begin{align}\label{eq:gabj_def}
U_{abj}(L)&=\exp \Big[L\gamma_{abj}\Big]\,.
\end{align}
Explicitly, we have, for the in-in case
\begin{align}
\label{eq:NGL_anom_dim}\gamma_{abj}&=-\log \Big(\cos \theta_j\Big)-\frac{1}{2}\log \Big(\frac{[ab]}{2[aj][jb]}\Big) \nonumber \\
&=-\log \Big(1-\tan ^2\frac{\theta_j}{2}\Big)-\log \Big(\frac{1+\cos \theta_j}{2}\Big)-\frac{1}{2}\log \Big(\frac{[ab]}{2[aj][jb]}\Big)\,.
\end{align}

\section{Collinear Resummation of Angle to Edge of Jet}\label{app:ci_of_U}

In this appendix we briefly summarize the kernels and scales appearing in the edge of jet factorization theorem discussed in \Sec{sec:coll_effect}. More details can be found in \Ref{Neill:2015nya}.

The inclusion of the edge-of-jet and boundary soft modes effects the next-leading logarithmic evolution. 
In the large-$N_C$ limit, and transforming to Laplace space, this amounts to modifying the $U$ evolution factor as 
\begin{align}
U_{abj}\Big(\tau_H^{-1},\tau_L^{-1}\Big)&\rightarrow U^{c.i.}_{abj}\Big(\tau_H^{-1},\tau_L^{-1}\Big)\,,
\end{align}
with the resulting collinearly improved BMS equation
\begin{align}
\tau_H\partial_{\tau_H}g_{ab}&=\frac{\alpha_s(\tau^{-1}_H)C_A}{\pi}\int_{J}\frac{d\Omega_j}{4\pi}W_{ab}(j)\Bigg(U^{c.i.}_{abj}\Big(\tau_H^{-1},\tau_L^{-1}\Big)g_{aj}g_{jb}-g_{ab}\Bigg)\,.
\end{align} 
If we write the collinearly improved resummation factor with arbitrary endpoints for the renormalization group, we have
\begin{align}
U^{c.i.}_{abj}\Big(\mu_{BS},\mu_{Dip.},\mu_{E.o.J.};\mu_i;\tau_H,\tau_L\Big)&=\nn \\
&\hspace{-3cm}U^{BS}_{nj}\Big(\mu_{BS},\mu_i\;\tau_H\Big)U_{abj}^{Dip.}\Big(\mu_{Dip.},\mu_i;\tau_L\Big)U^{\mathcal{E}}_{nj}\Big(\mu_{E.o.J.},\mu_i;\tau_L\Big)\,.
\end{align}
The anomalous dimension for each factor is written as follows
\begin{align}
\mu\frac{d}{d\mu}\text{ ln }U^{BS}_{nj}\Big(\mu,\mu_i\;\tau_H\Big)&=\gamma^{\Delta\theta}(\mu,\tau_H)\,,\\
\mu\frac{d}{d\mu}\text{ ln }U^{\mathcal{E}}_{nj}\Big(\mu,\mu_i\;\tau_L\Big)&=-\gamma^{\Delta\theta}(\mu,\tau_L)\,, \\
\label{eq:subtracted_dip_soft}\mu\frac{d}{d\mu}\text{ ln }U^{Dip.}_{abj}\Big(\mu,\mu_i\;\tau_L\Big)&=-4C_A\Gamma_{cusp}[\alpha_s(\mu)]\gamma_{abj}+\gamma^{\Delta\theta}(\mu,\tau_L)\,.
\end{align}
The anomalous dimension $\gamma_{abj}$ is defined in \Eq{eq:gabj_def}, and 
\begin{align}\label{eq:gamma_dtheta}
\gamma^{\Delta\theta}(\mu,\tau)&=4C_A\Gamma_{cusp}[\alpha_s(\mu)]\text{ln}\Bigg(\frac{e^{\gamma_E}\mu\tau\text{tan}\frac{R}{2}}{\text{tan}^2\frac{R}{2}-\frac{n\cdot n_q}{\nbar\cdot n_q}}\Bigg)\,.
\end{align}
Here we have made the assumption that we have a conical jet region of radius $R$.
The evolution factors result from evolving the boundary soft function, edge-of-jet soft function, and the subtracted dipole soft function respectively. The $W_{ab}(j)$ hard kernel in the BMS equation also has its own evolution equation that it implicitly obeys, but with appropriate scale choices, the logarithms can be taken to be minimized. From the explicit calculations of \Ref{Neill:2015nya}, the appropriate scales are found to be
\begin{align}
\mu_i&=\frac{\text{tan}^2\frac{R}{2}-\frac{n\cdot n_j}{\nbar\cdot n_j}}{\tau_He^{\gamma_E}\text{tan}\frac{R}{2}}e^{\gamma_{abj}} \,,\\
\mu_{Dip.}&=\frac{\text{tan}^2\frac{R}{2}-\frac{n\cdot n_j}{\nbar\cdot n_j}}{\tau_Le^{\gamma_E}\text{tan}\frac{R}{2}}e^{\gamma_{abj}} \,, \\
\mu_{BS}&=\frac{\text{tan}^2\frac{R}{2}-\frac{n\cdot n_j}{\nbar\cdot n_j}}{\tau_He^{\gamma_E}\text{tan}\frac{R}{2}}\,, \\
\mu_{E.o.J.}&=\frac{\text{tan}^2\frac{R}{2}-\frac{n\cdot n_j}{\nbar\cdot n_j}}{\tau_Le^{\gamma_E}\text{tan}\frac{R}{2}}\,.
\end{align}
This assumes we have started the running at the common scale chosen to be the scale of hard kernel of the BMS equation, with the boundary softs subtracted.\footnote{In \Ref{Neill:2015nya}, the boundary softs were not subtracted from the hard kernel, due to the form of the NLO kernel written in \Ref{Caron-Huot:2015bja}.} Note that the scale choice $\mu_{Dip.}$ is precisely where the anomalous dimension vanishes. For compactness sake, when we have made these canonical scale choices, we write:
\begin{align}
U^{c.i.}_{abj}\Big(\tau_H^{-1},\tau_L^{-1}\Big)&=U^{c.i.}_{abj}\Big(\mu_{BS},\mu_{Dip.},\mu_{E.o.J.};\mu_i;\tau_H,\tau_L\Big)\,.
\end{align}
We finally note that to mixed leading logarithmic order, we can swap $\tau e^{\gamma_E}\leftrightarrow m^{-1}$ to arrive at the resummation in cumulant space.

\bibliography{dressedgluon}

\end{document}